\documentclass[aps,prb,twocolumn,superscriptaddress,notitlepage]{revtex4-2}

\usepackage{amsmath,amssymb,mathtools,physics,graphicx,bm,hyperref}
\hypersetup{colorlinks=true,linkcolor=blue,citecolor=blue,urlcolor=blue}

\usepackage{comment}
\def\cred{\color{red}}

\newcommand{\ii}{\mathrm{i}}

\newcommand{\WY}[1]{\textcolor{blue}{[WY: #1]}}

\newcommand{\br}{\mathbf{r}}
\newcommand{\bK}{\mathbf{K}}
\newcommand{\bR}{\mathbf{R}}
\newcommand{\dg}{\dagger}
\newcommand{\om}{\omega}
\newcommand{\gm}{\gamma}
\newcommand{\al}{\alpha}

\newcommand{\ep}{\epsilon}
\newcommand{\dt}{\delta}

\begin{document}

\title{Majorana Signatures in Planar Tunneling through a Kitaev Spin Liquid}

\author{Weiyao Li}
\author{Vitor Dantas}
\affiliation{School of Physics and Astronomy, University of Minnesota, Minneapolis, MN 55455, USA}

\author{Wen-Han Kao}
\affiliation{Department of Physics, University of Wisconsin--Madison, Madison, WI 53706, USA}

\author{Natalia B. Perkins}
\affiliation{School of Physics and Astronomy, University of Minnesota, Minneapolis, MN 55455, USA}

\date{\today}

\date{\today}

\date{\today}

\begin{abstract}
We propose a planar tunneling setup to probe vacancy-bound Majorana modes in the chiral Kitaev spin liquid. In this geometry, the inelastic tunneling conductance can be expressed directly in terms of real-space spin correlations, establishing a  link between measurable spectra and the underlying fractionalized excitations. We show that spin vacancies host localized Majorana states that generate sharp near-zero-bias features, well separated from the continuum of bulk spin excitations. Compared to local STM measurements, the planar configuration naturally enhances the signal by coherently summing over multiple vacancies, reducing spatial resolution requirements. Our results demonstrate a realistic and scalable route to detect Majorana excitations in Kitaev materials.
\end{abstract}

\maketitle

\section{Introduction}

Quantum spin liquids (QSLs) arise in magnets where strong electronic correlations, spin–orbit–entangled exchange interactions, or geometrical frustration suppress conventional ordering tendencies and prevent long-range magnetic order even at zero temperature \cite{anderson1973resonating,Kitaev2006,Balents2010,Savary2016,KnolleMoessner2019,Takagi2019,Broholm2020}.
Through the fractionalization of spins into emergent gauge fields and unconventional quasiparticles such as spinons, Majorana fermions, and visons, QSLs manifest key concepts of lattice gauge theory including emergence, confinement, and topology \cite{Kogut1979}. A central challenge, however, is to understand the dynamics of these fractionalized excitations: how they propagate, interact, and couple to experimental probes. Understanding their dynamical signatures remains a major objective in the study of QSLs.

In Kitaev QSLs \cite{Winter2017,Hermanns2018,Motome2019,Trebst2022,Nasu2023, MatsudaRMP2025}, fractionalization takes an especially concrete form: each spin decomposes into itinerant Majorana fermions coupled to a static 
$\mathbb{Z}_2$ gauge field, giving rise to both extended bulk modes and localized flux excitations \cite{Kitaev2006}. Vacancies in the Kitaev model are known to trap localized Majorana modes whose dynamics imprint sharp, characteristic structures in spectroscopic observables \cite{Willans2010,Willans2011,Knolle2019,Nasu2020,Nasu2021,Kao2021vacancy,Kao2021localization,Dantas2022, Kao2024a, Kao2024b, Takahashi2023,Imamura2024,Yatsuta2024,Dantas2024}. When time-reversal symmetry is broken and the Kitaev QSL enters the non-Abelian phase \cite{Kitaev2006}, a vacancy induces an entire set of quasi-zero-energy in-gap Majorana modes from which a single protected zero mode ultimately emerges~\cite{Kao2024a,Kao2024b}. The localized nature of these vacancy-induced modes, together with their well-defined flux configuration, makes them particularly amenable to direct detection and controlled manipulation, which is the essential ingredient in several proposals for Majorana-based quantum computation. These properties make vacancies natural targets for local spectroscopic probes. 

\begin{figure*}[t]
  \centering
  \includegraphics[width=1\textwidth]{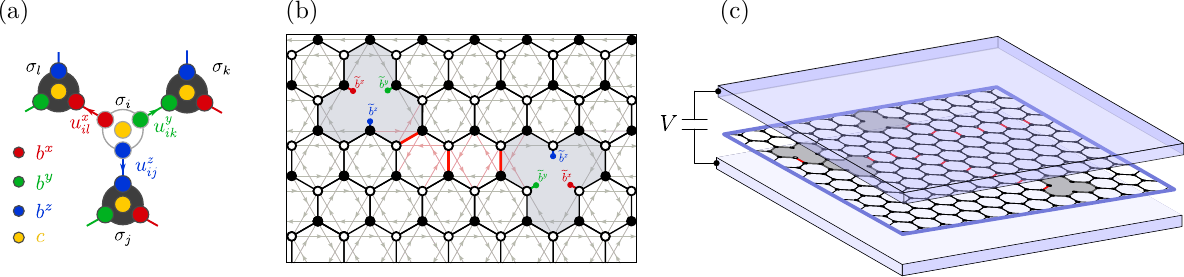}
  \caption{ (a) Illustration of spin fractionalization into Majorana fermions. 
(b) Kitaev honeycomb model with vacancies. Black and red bonds denote link variables 
$u_{ij}=+1$ and $u_{ij}=-1$, respectively. Gray-shaded plaquettes indicate 
$W_v=-1$, marking a $\pi$-flux bound to each vacancy. Three dangling Majorana modes 
$\tilde{b}^{\gamma}_j$ associated with each vacancy are also shown. 
(c) Schematic of the planar tunneling geometry: a substrate/Kitaev-QSL/substrate stack under bias $V$. 
The IETS signal $d^{2}I/dV^{2}$ directly probes the dynamical spin correlations 
$\sum_{jk} S^{\alpha\alpha}_{jk}(\omega)$. 
}
  \label{fig:fig1}
\end{figure*}

Scanning tunneling microscopy (STM) offers a direct and highly sensitive probe of these defect-bound Majorana modes \cite{konig2020tunneling,Feldmeier2020,Udagawa2021,Bauer2023,Takahashi2023,Kao2024a,Kao2024b,Peri2024, Kohsaka2024, Bauer2024, Jahin2025, Zhang2025PRB, Zhang2025npj}. Unlike bulk spectroscopies, STM can isolate the local electronic response of individual vacancies, enabling a real-space view of fractionalization at the atomic scale. Recent theoretical work has shown that vacancies in Kitaev QSLs generate pronounced features in the tunneling spectrum, including near-zero-frequency peaks stemming from localized Majorana modes and additional resonances arising from their hybridization with the itinerant Majorana continuum~\cite{Kao2024a,Kao2024b}.  
In particular, these studies identified several key signatures in the STM response that originate from the quasi-zero-energy modes. Most notably, they revealed a pronounced near-zero-bias peak in the derivative of the tunneling conductance whose position and intensity both increase with vacancy concentration, a behavior that directly reflects the fractionalized nature of these modes. These works also showed that STM can effectively probe the single-particle density of states of the Majorana fermions in the non-Abelian Kitaev QSL, providing a powerful route to accessing their low-energy dynamics.

However, STM-based detection remains technically challenging, as resolving these subtle low-energy structures requires exceptional tip stability and precise control over surface quality and defect environments.
In this work we build on the tunneling-junction proposal of König \emph{et al.} \cite{konig2020tunneling} for a Kitaev QSL barrier, and extend it to the scenario in which the spin liquid hosts dilute vacancies. Specifically, we propose a planar-junction geometry in which a vacancy-doped Kitaev QSL layer is sandwiched between two metallic electrodes. With an applied bias, \emph{inelastic electron tunneling spectroscopy} (IETS) probes internal spin correlations through the second derivative of the tunneling current, $d^{2}I/dV^{2}$.

This strategy is conceptually similar to the approach of Klein \emph{et al.}~\cite{Klein2018}, who
demonstrated that few-layer CrI$_3$ can act as a magnetic tunnel barrier, with IETS resolving magnon
excitations in the tunnel spectrum. By analogy, a vacancy-doped Kitaev spin-liquid layer embedded in
a planar junction is expected to exhibit distinctive in-gap signatures of fractionalized Majorana
excitations—most notably near-zero-bias peaks originating inside the bulk gap. Because $d^{2}I/dV^{2}$
is directly proportional to the dynamical spin–spin correlation function, this planar geometry
provides a practical and scalable route for detecting emergent Majorana excitations in Kitaev
materials.

 Motivated by these considerations, we next  analyze Majorana signatures in planar tunneling geometries for vacancy-doped Kitaev spin liquids, as illustrated in Fig.~\ref{fig:fig1}. Our starting point is the Hamiltonian
 \begin{equation}
\mathcal{H}=\mathcal{H}_{\rm{Kitaev}} +\mathcal{H}_{\rm Anderson}, 
\end{equation}
 where $\mathcal{H}_{\rm Kitaev}$ describes the Kitaev layer with vacancies and $\mathcal{H}_{\rm Anderson}$ accounts for the Anderson-type Hamiltonian that contains the free-electron Hamiltonian in the metallic leads and their coupling to the Kitaev spin liquid. Section~\ref{sec:model} introduces the vacancy-modified Kitaev honeycomb model. Section~\ref{sec:tunneling formalism} develops the planar-junction tunneling formalism and derives the inelastic current in terms of dynamical spin–spin correlation functions. Section~\ref{sec:spin_correlations} details the computation of these correlation functions using both analytical and numerical methods. Section~\ref{sec:results} presents the resulting tunneling spectra and demonstrates that the proposed planar geometry enables the detection of fractionalization in the Kitaev QSL. Additional derivations and supporting material are provided in the Appendices.

\section{Kitaev Honeycomb model}\label{sec:model}

% ---- optional figure stub

% \begin{figure}[t]
%   \centering
%   \includegraphics[width=0.9\columnwidth]{images/Kitaev Homeycomb Lattice Sketch_Final Paper One_Page 1.png}
%   \caption{Sketch of the Kitaev honeycomb model. The $\{x, y, z\}$-bond convention is illustrated in the bottom-left inset. Arrows in red mark bonds whose signs are flipped. Gray-shaded plaquettes correspond to eigenvalues $-1$, indicating a $\pi$-flux. Two true vacancies are outlined with dashed boundaries, with their corresponding enlarged plaquettes shown carrying bound flux. Three dangling $\tilde{b}^{\gamma}_j$ are presented for each vacancy. A pair of fluxes separated by a lattice distance $d$ is also shown.}
%   \label{fig:cleanS}
% \end{figure}

To model the non-Abelian Kitaev spin liquid in the presence of dilute spin vacancies, we consider the Hamiltonian
\begin{equation}
\mathcal{H}_{\rm Kitaev}=-J\sum_{\langle jk\rangle_\gamma}  \sigma^\gamma_j \sigma^\gamma_k-\kappa \sum_{\langle jkl\rangle_{\alpha\beta\gamma}}  
\sigma^\alpha_j \sigma^\beta_k \sigma^\gamma_l-h \sum_{j\in \mathbb{D}_\alpha}\tilde{\sigma}_{j}^{\alpha}.
\label{eq:HK}
\end{equation}
 The first term contains the anisotropic $\gamma=x,y,z$-type Ising interaction on corresponding nearest-neighbor bonds $\langle jk\rangle_{\gamma}$. The second is the time-reversal symmetry (TRS) breaking three-spin term of strength $\kappa$ 
 that opens a bulk gap $\Delta_{\rm bulk}$ and mimics the effect of a perturbative magnetic field. In the presence of vacancies, an external field of strength $h$  directly couples to a dangling spin component $\tilde{\sigma}^{\alpha}_j$  of site $j$ adjacent to vacancy sites, whose coordinates are collected in 
 $\mathbb{D}_{\alpha}$. For clarity, these dangling spin components are denoted with a tilde.

Using Kitaev’s representation of spins in terms of four Majoranas with $\sigma^\gamma_j=\ii b^\gamma_j c_j$ and introducing the link operators $
u^\gamma_{jk}\equiv \ii b^\gamma_j b^\gamma_k$, illustrated in Fig.~\ref{fig:fig1}(a), we obtain the Hamiltonian
\begin{align}
\mathcal{H}_{\rm Kitaev}&= \ii \sum_{\langle jk\rangle_\gamma} J_\gamma\, u^\gamma_{jk}\, c_j c_k \nonumber\\
&+\ii\kappa \!\!\sum_{\langle jkl\rangle_{\alpha\beta\gamma}}\!\!
u^\alpha_{jk} u^\gamma_{kl}\, c_j c_l-\ii h\sum_{j\in \mathbb{D}_\alpha} \tilde{b}_{j}^{\alpha} c_j,
\label{eq:HK_majorana}
\end{align}
where $\tilde{b}_j^{\alpha}$ denotes the dangling Majorana fermion arising from the decomposition of the dangling spin component $\tilde{\sigma}_j^{\alpha}$.
In this formulation, the interacting spin model factorizes into a static $\mathbb{Z}_2$ gauge sector, encoded in the link variables $\{u_{jk}\}$, and an itinerant Majorana $c$ sector. For any fixed gauge configuration $\{u_{jk}\}$, the Hamiltonian reduces to a quadratic Majorana problem.  Flux excitations correspond to changes in the eigenvalues of the plaquette operators
$W_p=\prod_{(jk)\in p}u_{jk}$, with $W_p=-1$ ($+1$) corresponding to  the presence (absence) of a vison on plaquette $p$.

In the absence of vacancies, the ground state of the Kitaev model resides in the zero-flux sector, as dictated by Lieb’s theorem~\cite{Lieb1994,Kitaev2006}.
Introducing a defect such as spin vacancy or impurity qualitatively alters this structure. 
Willans \textit{et al.}~\cite{Willans2010,Willans2011} showed analytically that, 
in the gapped phase of the Kitaev model with anisotropic  $(J_x, J_y, J_z)$, 
a single vacancy binds a $\mathbb{Z}_2$ flux in its vicinity. 
In the gapless phase near  the isotropic point $J_x$=$J_y$=$J_z$, this flux-binding effect persists and has been confirmed numerically~\cite{Willans2010,Willans2011,Kao2021vacancy,Dantas2022, ZschockeVojta2015}.
The three-spin interaction $\kappa$ further modifies the energetics of this bound-flux configuration: for small $\kappa$ the vacancy-bound flux remains favored, while beyond a critical $\kappa$ the zero-flux state becomes energetically preferable~\cite{Kao2021vacancy}. As a result, removing a spin reshapes the local gauge configuration by enlarging the adjacent plaquette [Fig.~\ref{fig:fig1}(b)], producing a characteristic bound-flux environment that strongly affects the low-energy Majorana spectrum. Similar flux-binding effects have been discussed in the Kitaev spin liquid in the presence of a Kondo impurity or a substitutional higher-spin impurity~\cite{Vojta2016impurity,Takahashi2025higherS}.

\begin{figure}[t]
  \centering
  \includegraphics[width=1\columnwidth]{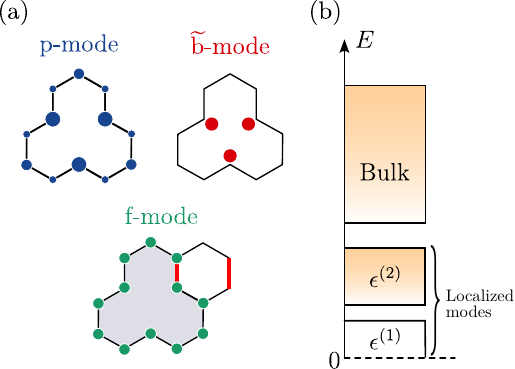}
  \caption{Different types of localized Majorana modes induced by a vacancy. 
The peripheral ($p$) mode has finite weight on the sites of the enlarged vacancy plaquette. 
The flux-induced ($f$) mode, present in the non-Abelian Kitaev spin liquid, is localized near the vacancy-bound $\mathbb{Z}_2$ flux. 
The $\tilde b$ modes originate from the fractionalization of dangling spin components adjacent to the vacancy. 
(b) Schematic energy spectrum of the quadratic Majorana problem, illustrating the hierarchy of localized modes involving itinerant $c$ fermions and dangling $\tilde b$ modes. 
}
  \label{fig:fig2}
\end{figure}

From the viewpoint of elementary localized modes, a vacancy generates a small set of degrees of freedom  [see Fig.~\ref{fig:fig2} (a)] whose hybridization structure governs the low-energy physics [see Fig.~\ref{fig:fig2} (b)]. In particular, the removal of a spin leaves three neighboring $\tilde b^\gamma$ Majorana operators unpaired, giving rise to three dangling $\tilde b^\gamma$ modes localized near the vacancy~\cite{Willans2011,Kao2024b}. In addition, the bound-flux configuration supports a peripheral ($p$) mode associated with the enlarged plaquette and a flux-induced ($f$) mode tied to the localized vison. In the presence of time-reversal-symmetry breaking, the $p$ and $f$ modes hybridize strongly with the itinerant $c$ Majoranas and are pushed to energies of order the bulk gap~\cite{Kao2024b}. By contrast, the dangling $\tilde b^\gamma$ modes hybridize only weakly and remain at parametrically low energies ($\epsilon^{(1)}$ and $\epsilon^{(2)}$). 
Consequently, these near-zero-energy $\tilde b$ modes dominate the vacancy-induced low-energy spin dynamics and form the primary focus of our analysis below.

% ---- optional figure stub

% \begin{figure}[t]
%   \centering
%   \includegraphics[width=0.9\columnwidth]{images/Kitaev Homeycomb Lattice Sketch_Final Paper One_Page 1.png}
%   \caption{Sketch of the Kitaev honeycomb model. The $\{x, y, z\}$-bond convention is illustrated in the bottom-left inset. Arrows in red mark bonds whose signs are flipped. Gray-shaded plaquettes correspond to eigenvalues $-1$, indicating a $\pi$-flux. Two true vacancies are outlined with dashed boundaries, with their corresponding enlarged plaquettes shown carrying bound flux. Three dangling $\tilde{b}^{\gamma}_j$ are presented for each vacancy. A pair of fluxes separated by a lattice distance $d$ is also shown.}
%   \label{fig:cleanS}
% \end{figure}

% ==================== Section IV ====================
\section{Planar Tunneling Formalism}\label{sec:tunneling formalism}
In this section we focus on the tunneling term of the Hamiltonian, $\mathcal{H}_{\rm tunnel}$.
Our aim is to derive an explicit expression for how electrons in the metallic electrodes co-tunnel through the Kitaev spin-liquid barrier, and how its  spin dynamics becomes imprinted onto the resulting inelastic current.
 As illustrated in Fig.~\ref{fig:fig1}(c), a single Kitaev QSL layer is sandwiched between two metallic leads under an applied bias $V$. Following Ref.~\cite{konig2020tunneling}, we  first outline the effective low-energy tunneling Hamiltonian and explain how it determines the inelastic conductance.

\subsection{Effective tunneling Hamiltonian}

%(\WHK{Shall we consider making $\hat{\sigma} \to \sigma$ in this and next section (and the Appendix) so that the Pauli spin operator has the same notation as in the Kitaev Hamiltonian section? Or any reason to make a distinction?})

Electrons in the metallic leads can tunnel through the Kitaev spin-liquid layer by coupling to the sites at the interface.
At each interface site ${\bf r}$, the electron degrees of freedom in the Kitaev QSL are represented by a localized orbital 
$d_{{\bf r}}$ that  locally hybridizes with the substrate fermions  $\check{c}_{{\bf r}\sigma \xi}$ with spin indices $\sigma=\uparrow,\downarrow$ and substrates label $\xi=1,2$. 
The corresponding Anderson-type Hamiltonian reads
\begin{align}
\mathcal{H}_{\text{Anderson}}&=\sum_{{\bf r}}\big(E_d d_{{\bf r}}^{\dagger} d_{{\bf r}}+U d^{\dagger}_{{\bf r}\uparrow} d_{{\bf r}\uparrow}d^{\dagger}_{{\bf r}\downarrow} d_{{\bf r}\downarrow}\big)\nonumber\\
&+\sum_{\mathbf{k},\xi,\sigma}\varepsilon_{\mathbf{k}}\, \check{c}^\dagger_{\mathbf{k}\sigma\xi}\check{c}_{\mathbf{k}\sigma\xi}
+\sum_{{\bf r},\sigma,\xi}\!\big(g_{\bf r} \check{c}^\dagger_{{\bf r}\sigma\xi}d_{{\bf r}\sigma}+\text{H.c.}\big),
\label{eq:anderson}
\end{align}
where 
$E_{d}$ denotes the  on-site energy  and 
$U$ its on-site Coulomb repulsion on the localized $d$-orbital, $\varepsilon_{k}$ the dispersion of 
electrons in the metallic leads, and $g_{{\bf r}}$ is the local hybridization 
amplitude at interface site ${\bf r}$. In the dilute-vacancy regime considered 
here, any missing sites have $g_{{\bf r}}=0$.

At energies well below the charge–excitation scale of the localized orbital,
set by $E_{d}$ and $U$, the charge fluctuations can be integrated out by using the
Schrieffer–Wolff (SW) transformation. This yields a low-energy tunneling 
Hamiltonian of Kondo type, in which virtual charge excitations generate 
effective spin-exchange processes between the substrate electrons and the 
spin degrees of freedom of the Kitaev layer:
\begin{equation}
\mathcal{H}_{\rm tunnel}=\sum_{{\bf r},\sigma,\sigma^{\prime}}\sum_{\xi, {\xi}^{\prime}} \check{c}^\dagger_{{\bf r}\sigma \xi}\left[\,\mathcal{J}_{\bf r}\,\boldsymbol{\tau}_{\sigma\sigma'}\!\cdot\!\boldsymbol \sigma_{\bf r}
+t_{\bf r}\delta_{\sigma\sigma'}\right]\check{c}_{{\bf r}\sigma^{\prime} {\xi}^{\prime}},
\label{eq:Htun}
\end{equation}
where $\bm{\tau}$ 
are the Pauli matrices acting on the spin of the 
substrate electrons, and $\boldsymbol{\sigma}_{\bf r}$ denotes the local spin 
operator of the Kitaev layer at site ${\bf r}$. The SW transformation generates the 
effective couplings
\begin{align}
\mathcal{J}_{\bf r} &=\frac{|g_{\bf r}|^2}{2}\!\left(\frac{1}{U+E_d}-\frac{1}{E_d}\right),\\
t_{\bf r} &=-\frac{|g_{\bf r}|^2}{2}\!\left(\frac{1}{U+E_d}+\frac{1}{E_d}\right),
\label{eq:Jt}
\end{align}
 where $\mathcal{J}_{\bf r}$ governs  inelastic spin-flip co-tunneling processes that 
directly probe the spin dynamics of the Kitaev layer, while $t_{\bf r}$ describes 
elastic potential scattering and contributes only to a background component of 
the conductance.
A step-by-step derivation of the effective exchange and potential-scattering amplitudes is presented in the Appendix~\ref{APP:A1}.

 \subsection{Inelastic current}\label{subsec:inelastic}

With an applied bias $V$ across the planar junction, electrons can tunnel
inelastically by exchanging energy with the spin degrees of freedom in the
Kitaev layer.  Using the exchange part of the effective tunneling Hamiltonian
[Eq. \eqref{eq:Htun}], Kubo’s formula yields 
\begin{eqnarray}
I_{\rm inel}(V)
&=&
4\sum_{{\bf r},{\bf r}'}\int_0^{eV}\!\frac{d\omega}{2\pi}\,
\mathcal{J}_{\bf r}\mathcal{J}_{{\bf r}'}\,
\Im C^{R}({\bf r},{\bf r}';\omega)
\nonumber\\
&\times&
\Big[\Im \Pi^{R}_{12}(eV-\omega)
     -\Im \Pi^{R}_{21}(\omega-eV)\Big],
\label{eq:Iinel_general_main}
\end{eqnarray}
where $C^R$ is the retarded spin correlator of the Kitaev layer,
\begin{eqnarray}
C^R({\bf r},{\bf r}';\omega)
&=&
-i\!\int_0^\infty\! dt\, e^{i\omega t}\,
\langle[\sigma_{\bf r}(t),\sigma_{{\bf r}'}(0)]\rangle_{\rm QSL},
\label{eq:CR_main}
\end{eqnarray}
and $\Pi^R_{\xi\xi'}$ is the  retarded spin bubble    of the metallic leads,
\begin{eqnarray}
\Pi^R_{\xi\xi'}({\bf r}-{\bf r}';\omega)
&=&
-i\!\int_0^\infty\! dt\, e^{i\omega t}
\nonumber\\
&\times&
\langle[
\check{\mathbf c}^\dagger_{{\bf r}\xi}(t)\!\cdot\!\boldsymbol{\tau},
\check{\mathbf c}_{{\bf r}'\xi'}(0)\!\cdot\!\boldsymbol{\tau}
]\rangle_{\rm lead}.
\label{eq:PiR_main}
\end{eqnarray}
The microscopic evaluation of $\Pi^R$, is provided in Appendix~\ref{APP:A2}.  Its essential
consequence is that for a uniform two–dimensional metal,
$\Im \Pi^R \propto |\omega|$, allowing the inelastic current to be written
directly in terms of the Kitaev spin response.

For a uniform interface, except at vacancy sites $\mathbb{V}$ where
$\mathcal{J}_{\bf r}=0$, the exchange couplings reduce to
$\mathcal{J}_{\bf r}=\mathcal{J}_0$ for all ${\bf r}\notin\mathbb{V}$.  Eq.~(\ref{eq:Iinel_general_main}) therefore simplifies to
\begin{eqnarray}
\frac{d^2 I_{\rm inel}}{dV^2}
&=&
-8\left(\frac{\nu_{2\rm D}q_{\max}}{2\pi v_F}\right)^{2}
\mathcal{J}_0^2
\sum_{{\bf r},{\bf r}'\notin\mathbb{V}}
\Im C^R({\bf r},{\bf r}';\omega=eV),\nonumber\\
\label{eq:d2I_main_eqnarray}
\end{eqnarray}
where $\nu_{2\rm D}$ is the two–dimensional density of states, $v_F$ is the
Fermi velocity of the metallic substrate, and $q_{\max}$  is  an ultraviolet cutoff.

At zero temperature,
\begin{eqnarray}
\Im C^{R}({\bf r},{\bf r}';\omega)
&=&
-\frac12\int_{-\infty}^{\infty} dt\,
e^{i\omega t}\,
\langle\sigma_{\bf r}(t)\sigma_{{\bf r}'}(0)\rangle_{c},
\end{eqnarray}
where $\langle \mathcal{A}\mathcal{B}\rangle_{c}
=\langle \mathcal{A}\mathcal{B}\rangle-\langle \mathcal{A}\rangle\langle \mathcal{B}\rangle$ denotes the connected
correlator.  Defining the dynamical spin structure factor
$S^{\alpha\alpha}({\bf r},{\bf r}';\omega)
=\int_{-\infty}^{\infty} dt\, e^{i\omega t}
\langle \sigma_{\bf r}^\alpha(t)\sigma_{{\bf r}'}^\alpha(0)\rangle_c$, we obtain
\begin{eqnarray}
\frac{d^2 I_{\rm inel}}{dV^2}
&\propto&
\sum_{\alpha}\sum_{{\bf r},{\bf r}'\notin\mathbb{V}}
S^{\alpha\alpha}({\bf r},{\bf r}';\omega=eV).
\label{eq:S_main}
\end{eqnarray}
Thus the second derivative of the inelastic current provides a direct,
spatially summed measure of the dynamical spin correlations of the Kitaev
spin liquid, linking the tunneling response to the system's fractionalized
spin dynamics.

\begin{figure*}[t]
\includegraphics[width=1\textwidth]{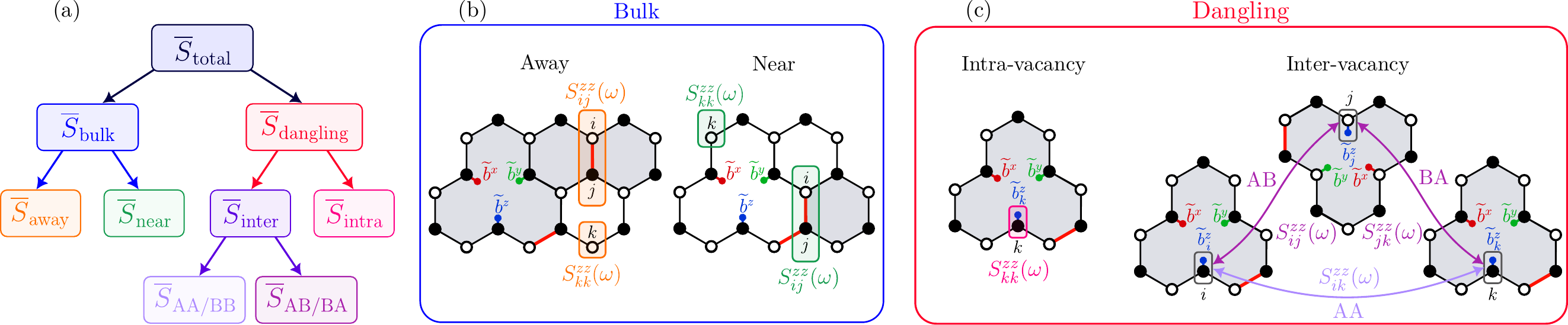}
\caption{(a) Tree diagram summarizing the decomposition of the full  dynamical spin spectral weight into different components contributing to the planar tunneling response. (b) Illustration of bulk spin–spin correlations, separated into contributions from bonds ``away" from vacancies (not sharing any bond with a vacancy; the closest such case is shown) and bonds 
``near"  a vacancy (in the immediate vicinity, sharing a bond with the vacancy). The near-vacancy component reflects locally modified flux gaps and produces weak in-gap spectral weight, while the away-vacancy contribution reproduces the bulk two-flux continuum. In both cases, we include on-site and nearest-neighbor contributions to $S_{ij}^{\alpha\alpha}(\omega)$. (c)
Schematic decomposition of  spin correlations involving  dangling spin components adjacent to vacancies into intra-vacancy and inter-vacancy components. Inter-vacancy contributions are further resolved into same-sublattice (AA/BB) and opposite-sublattice (AB/BA) channels, highlighting their distinct interference behavior in the planar tunneling response.} 
\label{fig:flow diagram}
\end{figure*}

\subsection{Hierarchy of contributions to the planar tunneling signal}
\label{subsec:hierarchy}

Since the planar tunneling response is proportional to a sum over all spin--spin correlation functions in the system, it is essential to understand how 
the different types of correlations contribute to the measurable signal. Unlike STM, which probes only the on-site and nearest-neighbor correlations beneath the tip, the planar geometry produces a spatially averaged and coherent response. As a result, the planar spectrum contains contributions with distinct physical origins: extended bulk correlations, localized correlations associated with vacancies, and non-local interference terms between vacancy-induced modes. Organizing these contributions in a systematic and physically transparent way is therefore crucial for identifying which features of the tunneling response arise from bulk Majorana dynamics and which originate from the fractionalized degrees of freedom bound to vacancies.

To facilitate this analysis, we introduce a decomposition of the total disorder-averaged dynamical spin-spin correlation function into contributions that reflect the microscopic structure of the system. A schematic representation of this organization is shown in Fig.~\ref{fig:flow diagram} (a).

We begin by separating the total spectral function into a  vacancy-induced or \emph{dangling} part and a \emph{bulk} part:
\begin{equation}
    \bar{S}_{\rm total}
    = \bar{S}_{\rm bulk}
    + \bar{S}_{\rm dangling},
\end{equation}
where the overline denotes disorder averaging.
The dangling sector includes all correlations involving  dangling spin components adjacent to vacancies and is responsible for the near-zero-bias structure discussed later. It naturally decomposes into intra- and inter-vacancy components [Fig.~\ref{fig:flow diagram} (c)]:
\begin{equation}
    \bar{S}_{\rm dangling}
    = \bar{S}_{\rm intra} + \bar{S}_{\rm inter}.
\end{equation}
$\bar{S}_{\rm Intra}$ collects correlations among sites adjacent to a single vacancy, while $\bar{S}_{\rm inter}$ captures correlations involving sites adjacent to different vacancies. The inter-vacancy term can be further resolved according to sublattice type of the sites involved,
\begin{equation}
    \bar{S}_{\rm inter}
    = \bar{S}_{\rm AA/BB}
    + \bar{S}_{\rm AB/BA},
\end{equation}
distinguishing same-sublattice (AA/BB) from opposite-sublattice (AB/BA) contributions.

On the other hand, the bulk sector includes all spin-spin correlations not associated with  dangling spin components.  For numerical clarity, we further partition the bulk response into near- and far-from-vacancy components because the response from these two regions behave differently in the presence of vacancy-bound fluxes:
\begin{equation}
    \bar{S}_{\rm bulk}
    = \bar{S}_{\rm{away}}
    + \bar{S}_{\rm{near}}.
\end{equation}
The main difference between $\bar{S}_{\rm away}$ and $\bar{S}_{\rm near}$ is that the former is associated to a two-flux excitation gap $\Delta_{2f}$ on the hexagonal plaquettes as in the pure Kitaev model, while the latter is associated to a smaller two-flux excitation gap $\tilde{\Delta}_{2f}$ on one hexagonal and one vacancy plaquette [Fig.~\ref{fig:flow diagram}~(b)].  Note that each contribution contains both the on-site and nearest-neighbor correlations for the bulk spins.

This hierarchical decomposition provides the structural framework for interpreting the planar tunneling spectra presented in Sec.~\ref{sec:results}. 

% ==================== Section III ====================

\section{Spin Dynamics}\label{sec:spin_correlations}
Having expressed the planar tunneling signal in terms of KSL spin correlation functions, we now turn to computing these quantities explicitly. We first recall their forms in the clean system, which establishes the selection rules, and then analyze how vacancies modify these correlations through localized dangling Majorana modes.

From this point on, for clarity of notation, we label lattice sites
by $j$ and denote their coordinates by $\mathbf{r}_j$,
 rather than labeling them directly by their coordinate vectors $\mathbf{r}$.

\subsection{Spin correlations in the pure Kitaev model}\label{subsec:clean}

Using the Majorana fractionalization $\sigma_{j}^{\gamma}=ib_{j}^{\gamma}c_j$ from Sec.~\ref{sec:model}, we can express each spin operator in terms of bond  $\chi_{\langle jk\rangle\alpha}=\tfrac{1}{2}(b^\alpha_j-\ii b^\alpha_k)$ and matter  fermions $c_j$ as
\begin{equation}
\sigma^\alpha_j=\eta_j\big(\chi^\dagger_{\langle jj'\rangle\alpha}+\zeta_j\chi_{\langle jj'\rangle\alpha}\big)c_j,
\label{eq:sigma_rewrite}
\end{equation}
where
\begin{equation}
(\eta,\zeta)=
\begin{cases}
(\ii, +1),& j\in A,\\
(1, -1),& j\in B.\\
\end{cases}
\end{equation}
Evaluating $S^{\alpha\beta}_{jk}(t)$ in the clean model ground state flux sector, where all plaquette operators carry eigenvalue $+1$ \cite{Knolle2014, Knolle2015Fractionalization}, yields
\begin{align}
{S}_{jk}^{\alpha\beta}(t) & =-\eta_{j}\eta_{k}\zeta_{j}\langle F|\otimes \langle M|e^{iH_{u_{0}}t}c_{j}(0)\chi_{\langle jk^{\prime}\rangle_{\alpha}}(0)\nonumber\\
&\ \ \ \ \times e^{-iH_{u_{l}}t}\chi_{\langle kj^{\prime}\rangle_{\beta}}^{\dagger}(0)c_{k}(0)|F\rangle \otimes |M\rangle\nonumber \\
 & =-\eta_{j}\eta_{k}\zeta_{j}\delta_{\alpha\beta}\delta_{\langle jk\rangle\alpha}\nonumber\\
 &\ \ \ \ \times \langle M|e^{iH_{u_{0}}t}c_{j}e^{-iH_{u_{l}}t}c_{k}|M\rangle.
 \label{eq:cleanCorr}
\end{align}
Here  the ground state  is written as  $|0\rangle\equiv|F\rangle\otimes|M\rangle$ to emphasize it has both flux and matter parts. $H_{u_0}, H_{u_l}$ represent the Kitaev Hamiltonian evaluated, respectively, in the ground-state flux sector and in the sector with a pair of locally excited fluxes. Because a spin operator  induces two fluxes in the adjacent plaquettes, the spin–spin correlation function probes only states above the two-flux excitation gap $\Delta_{2f}\approx0.26J$ \cite{Kitaev2006, Aaditya2023}.
Furthermore, in the absence of vacancies, only on-site ($j=k$) and nearest–neighbor (NN) pairs contribute to the spin-spin correlation function of the Kitaev model~\cite{Baskaran2007,Knolle2014,Knolle2015Fractionalization}.  Note that in our tree diagram [Fig.~\ref{fig:flow diagram}~(a)], the above contribution corresponds to $\bar{S}_{\rm away}$ because the two-flux excitation does not involve a vacancy plaquette.

\subsection{Spin correlations in the presence of vacancies}\label{subsec:vac_S}

In the presence of vacancies, translational invariance is broken and the local flux background is modified. As a result, the spin-spin correlations entering Eq.~\eqref{eq:S_main} acquire additional nonlocal contributions involving sites adjacent to vacancies. As shown in Eq.~\eqref{eq:S_main}, the inelastic tunneling signal probes spin-diagonal correlation functions $S^{\alpha\alpha}({\bf r},{\bf r}';\omega)$, accordingly, throughout this section we restrict attention to the diagonal components $S^{\alpha\alpha}_{jk}(\omega)$.

These new spin-spin correlations due to vacancies still involve the \emph{dangling} and \emph{bulk} contributions. The dangling contribution consists of $\bar{S}_{\rm inter}$ and $\bar{S}_{\rm intra}$, but both of them can be written as:
\begin{align}
\bar{S}_{\rm dangling}
&=\sum_{\alpha}\sum_{j,k\in\mathbb{D}_{\alpha}}
\bar{S}^{\alpha\alpha}_{jk},
\label{eqs:bulk_dangling2}
\end{align}
where $\mathbb{D}_{\alpha}$ denotes the set of sites adjacent to vacancies with the dangling spin components $\alpha$. The dangling sector collects correlations between pairs of dangling spin components and encodes the vacancy-induced low-energy physics.

For the bulk part involving sites adjacent to vacancies, it is precisely the $\bar{S}_{\rm near}$ in Fig.~\ref{fig:flow diagram}~(a):
\begin{align}
\bar{S}_{\rm near}
&=\sum_{\alpha}\sum_{j,k\notin\mathbb{D}_{\alpha}}
\bar{S}^{\alpha\alpha}_{jk}.
\label{eqs:bulk_dangling1}
\end{align}

The analytic structure of these contributions is most transparent when written in a Lehmann representation. In the following, we evaluate the dangling and bulk correlation functions by diagonalizing the quadratic Majorana Hamiltonian and expressing the resulting matrix elements in terms of Bogoliubov coefficients. For completeness, all expressions used in the numerical calculations are summarized in Appendix \ref{app:B}.

\begin{figure*}[t]
  \centering
  \includegraphics[
    width=\textwidth,
    height=0.82\textheight,
    keepaspectratio,
    page=1
  ]{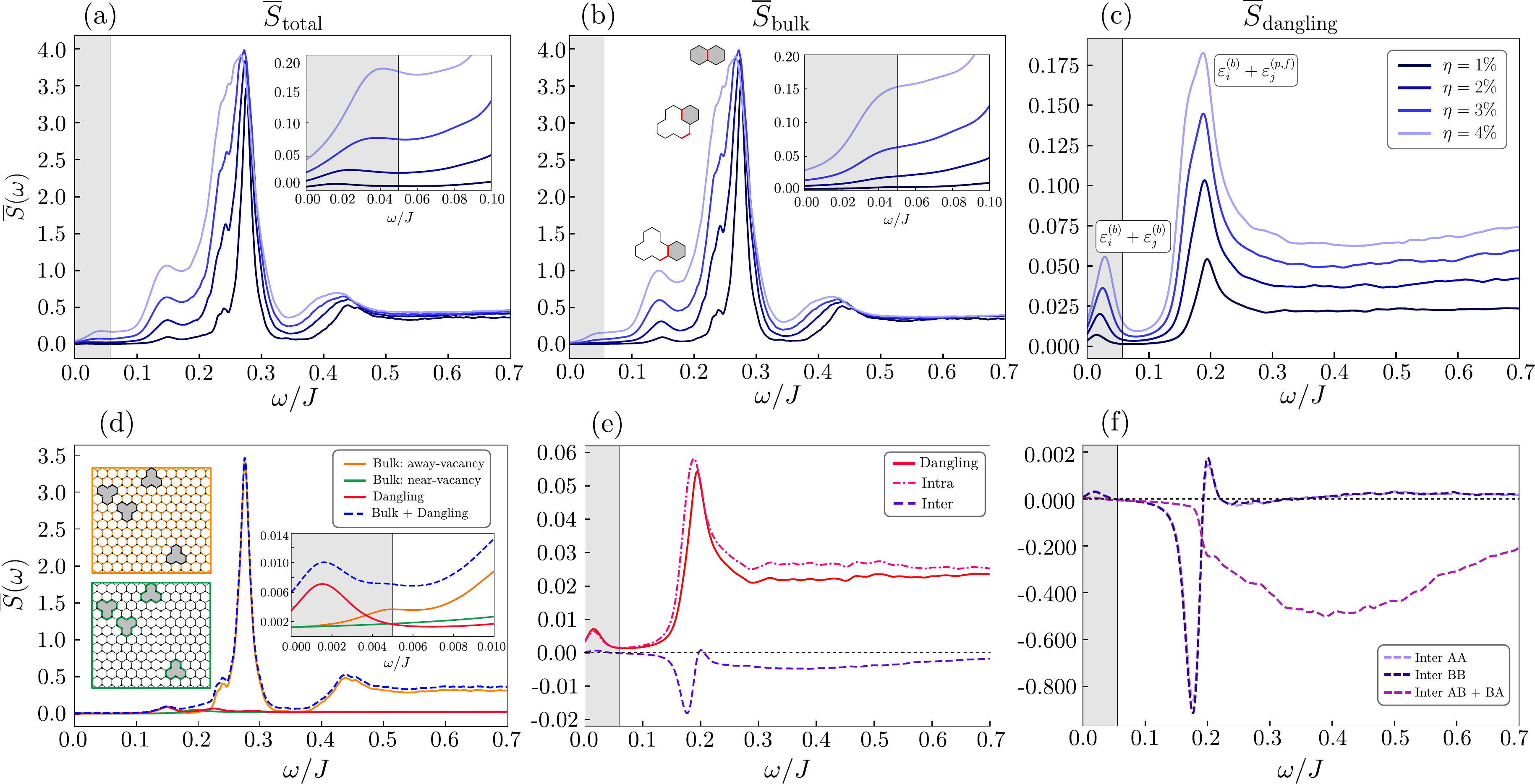}
  \caption{\label{fig:wide} 
   (a)–(c) Disorder-averaged spin–spin correlation functions  for different vacancy densities:
(a) total correlation function $\bar{S}_{\rm total}(\omega)$.
The inset shows a magnification of the near-zero-bias peak; its intensity is
approximately $1\%$ of the main peak near the bulk gap
$\Delta_{\rm bulk}\approx0.2J$ (dashed line).
(b) Bulk contribution $\bar{S}_{\rm bulk}(\omega)$, with the
corresponding flux excitation energies indicated.
(c) Dangling contribution $\bar{S}_{\rm dangling}(\omega)$.
The two dominant low-energy peaks arise from distinct combinations of
vacancy-induced low-energy modes \cite{noteVacancies}.
   (d)–(f) Comparison of different contributions at a fixed vacancy
density $\eta=1\%$:
(d) near-vacancy bulk, away-vacancy bulk, and dangling contributions.
Sketches illustrate near-vacancy (top) and away-vacancy (bottom) bulk
processes; the inset shows a zoom of the near-zero-bias region.
(e) intra-vacancy and inter-vacancy dangling contributions.
(f) Same-sublattice (AA/BB) and opposite-sublattice (AB+BA)
inter-vacancy dangling contributions. 
The shaded region in all panels highlights the near-zero-bias window of interest.
Parameters are $\kappa=0.02J$, $h=0.05J$, and Lorentzian broadening
$\delta=0.01$.  Each plot is individually normalized.
}
\end{figure*}

For the dangling sector, the zero-frequency contribution,
\begin{equation}
S^{\alpha\alpha\,(0)}_{jk}(\omega)=\delta(\omega)\,\langle \tilde{\sigma}^\alpha_j\rangle\langle \tilde{\sigma}^\alpha_k\rangle,
\label{eq:elastic}
\end{equation}
does not contribute to the inelastic response and is therefore omitted. The leading contribution arises from two-particle intermediate states,
\begin{align}
S^{\alpha\alpha\,(2)}_{jk}(\omega)&=
-\sum_{\gamma,\delta}\delta\!\big(\omega-\epsilon_\gamma-\epsilon_\delta\big)\nonumber\\
&\times
\langle \tilde{0}| \tilde{b}^\alpha_j c_j a^\dagger_\gamma a^\dagger_\delta |\tilde{0}\rangle
\langle \tilde{0}| a_\delta a_\gamma \tilde{b}^\alpha_k c_k |\tilde{0}\rangle,
\label{eq:dang2}
\end{align}
where $|\tilde{0}\rangle$ denotes the bound-flux sector, in which all plaquettes away from vacancies remain in the zero-flux configuration while $\mathbb{Z}_2$ fluxes are bound to vacancies. The energies $\epsilon_\gamma$ and $\epsilon_\delta$ correspond to the intermediate eigenstates. In the numerical evaluation, the matrix elements in Eq.~\eqref{eq:dang2}  are computed via a Bogoliubov transformation of the quadratic Majorana Hamiltonian, yielding closed-form expressions for all matrix elements (see Appendix~\ref{app:B}).

We now turn to the remaining bulk contributions. For non-dangling spin components, the spin-spin correlation function retains a structure analogous to that of the clean system, Eq.~\eqref{eq:cleanCorr}, but evaluated in the vacancy-induced bound-flux background. Retaining diagonal contributions only, we obtain
\begin{align}
\mathcal{S}_{jk}^{\alpha\alpha}(t)
&=
-\eta_{j}\eta_{k}\,
\delta_{\langle jk\rangle_\alpha}
\nonumber\\
&\quad\times
\Big\{
\zeta_{j}\,
\langle \tilde{0}|
e^{iH_{u_{0}}t}c_{j}
e^{-iH_{u_{l}}t}c_{k}
|\tilde{0}\rangle
\,\delta_{n_{jk},0}
\nonumber\\
&\qquad+
\zeta_{k}\,
\langle \tilde{0}|
e^{iH_{u_{0}}t}c_{j}
e^{-iH_{u_{l}}t}c_{k}
|\tilde{0}\rangle
\,\delta_{n_{jk},1}
\Big\},
\label{eq:bulk_correlation_expression}
\end{align}
where $n_{jk}$ denotes the occupation number of the bond fermion associated with the bond $\langle jk\rangle_\alpha$. Equivalently, $n_{jk}$ tracks the $\mathbb{Z}_2$ gauge link on that bond. The two terms correspond to the two possible bond-fermion parities: for $n_{jk}=0$ the contribution is weighted by $\zeta_j$, while for $n_{jk}=1$ it is weighted by $\zeta_k$ (introduced in Eq.~(\ref{eq:sigma_rewrite})), reflecting the change in fermionic sign structure induced by the flipped bond.
Compared to the clean case, the local two-flux excitation energy entering the time evolution is replaced by a position-dependent gap $\tilde{\Delta}_{2f}$ related to  the vacancy-induced bound-flux background \cite{Takahashi2023,Kao2024b}. Typically, $\tilde{\Delta}_{2f}$ is smaller than the clean two-flux gap $\Delta_{2f}$.

\section{Numerical results}\label{sec:results}

Our numerical calculations are done on a $40\times 40$ lattice with 100 
independent disorder realizations at dilute vacancy density $\eta$ ($1\%,2\%,3\%,4\%$). The model parameters are chosen as   $\kappa=0.02J$ and $h=0.05J$,  and a Lorentzian broadening factor $\delta=0.01$ is used throughout. To estimate the bulk contributions far from vacancies, $\bar{S}_{\rm away}$, we randomly sampled $500$ bulk bonds in each disorder realization and used their mean as  a representative background contribution.

The resulting disorder-averaged spin–spin correlation functions
$\bar S(\omega)$ are shown in Fig.~\ref{fig:wide}.  As discussed in
Sec.~\ref{sec:tunneling formalism} and summarized in
Eq.~(\ref{eq:S_main}), the experimentally measured inelastic tunneling
signal is
$d^2 I_{\rm inel}/dV^2 \propto \bar S(\omega=eV)$, so that the spectral
features in Fig.~\ref{fig:wide} directly translate into features of the
tunneling spectrum.

In Fig.~\ref{fig:wide} (a), we present the total spin–spin correlation
function $\bar{S}_{\rm total}(\omega)$ for several vacancy
densities. The spectrum is dominated by bulk spin excitations, except within the bulk gap where vacancy-induced contributions prevail.
The bulk response, $\bar{S}_{\rm bulk}(\omega)$, 
shown separately in Fig.~\ref{fig:wide}(b), forms a broad continuum with a prominent peak at the two-flux excitation
 $\Delta_{2f}\approx0.26J$, inherited from
the clean model and originated from the away-from-vacancy component
$\bar{S}_{\rm  away}$.
Vacancies also modify the low-energy part of the bulk spectrum. As illustrated in Fig.~\ref{fig:wide}(b), bulk correlations
involving sites near a vacancy,
$\bar{S}_{\rm  near}$, exhibit flux excitations at reduced
energies, reflecting the locally suppressed flux gap near a vacancy.

Superimposed on these continuum are weaker 
but clearly resolved
in-gap features arising
from vacancy-bound modes (see Fig.~\ref{fig:fig2}). Their origin is highlighted by isolating the dangling contribution,
shown separately in Fig.~\ref{fig:wide}(c).
The relative smallness of this contribution follows from a simple counting argument:
 in the planar geometry the number of bulk pairs scales as $(1-\eta)8N$ (three NN and one on-site per site, over $2N$ sites), whereas  the main near-zero-bias  contribution from intra-vacancy pairs scale as $6\eta N$. For $\eta\!=\!1\%$ this yields a $\sim\!1\%$ ratio, consistent with our numerical results. The near-zero-bias peaks occur at frequencies corresponding to sums of two low-energy
 $\tilde b$-like fermionic excitations associated with vacancies, as described by Eq.~(\ref{eq:dang2}),
 with intensities determined by the matrix elements of the spin operators between the ground state and the corresponding two-particle vacancy states. The systematic enhancement of the near-zero-bias signal with increasing vacancy density therefore reflects the growing number of vacancy-induced
 $\tilde b$-modes and provides direct evidence for spin fractionalization in the Kitaev spin liquid.

To further clarify the origin of the in-gap spectral weight, Fig.~\ref{fig:wide}(d) compares the near-vacancy bulk, away-vacancy bulk, and dangling contributions at   fixed vacancy density $\eta\!=\!1\%$.
The near-zero-bias response in Fig.~\ref{fig:wide}(d) is clearly dominated by the dangling contribution, demonstrating that the lowest-energy spectral features originate from localized vacancy-induced degrees of freedom rather than from bulk processes. 
This dominance persists despite the presence of subleading in-gap weight from near-vacancy bulk correlations, providing a controlled basis for interpreting the near-zero-bias signal as a direct manifestation of Majorana fractionalization.

\begin{comment}$\bar{S}_{\rm  near}$ also contribute spectral
weight inside the bulk gap [see Fig.~\ref{fig:wide}(b)]. This reflects
the fact that flux excitations involving sites near a vacancy are
energetically softened compared to the clean system, since flipping
fluxes in the vicinity of a vacancy costs less energy. 
In addition, nearly zero-energy
vacancy-induced peripheral ($p$) modes and, in the non-Abelian phase
($\kappa\neq0$), flux-induced ($f$) modes hybridize with the itinerant
Majorana fermions. This hybridization shifts these modes into the bulk
gap, where they appear as finite-energy in-gap resonances in the
near-vacancy bulk contribution $\bar{S}_{\rm  near}$.
\end{comment}

 In the intermediate in-gap energy range, $\omega/J \in (0.1,\,0.2)$, the spectral weight is dominated by bulk contributions, as is evident from the relative scales in Fig.~\ref{fig:wide}(b) and Fig.~\ref{fig:wide}(c).
 In this regime, the dangling contribution does not qualitatively affect the overall response. Nevertheless, within this subgap  energy window the dangling sector exhibits a nontrivial internal structure. As illustrated in Fig.~\ref{fig:wide}(e,f), the total dangling contribution can be further decomposed into intra- and inter-vacancy parts. Within the subgap energy range, the on-site (intra-vacancy) term $\bar{S}_{\mathrm{dangling}}(\omega)$ clearly dominates and remains strictly positive, which follows directly from the amplitude-squared structure of the single-site  component in the correlation function, Eq.~(\ref{eq:dang2}).
 Inter-vacancy dangling correlations, on the other hand, involve nonlocal processes and decay rapidly with vacancy separation. As shown in Appendix~\ref{app:sublattice}, their cumulative contribution therefore remains subleading in the thermodynamic limit, justifying the dominance of the intra-vacancy term in the dangling response.

Although parametrically smaller than the intra-vacancy contribution, the inter-vacancy correlations nevertheless encode nontrivial structure. Their influence is most clearly visible within the subgap window, where same-sublattice (AA/BB) correlations give rise to a characteristic dip-peak feature, as shown in Fig.~\ref{fig:wide}(e,f). This spectral feature originates from interference between vacancy-induced in-gap modes associated with different vacancies.

The same interference effects also govern the spatial dependence of the inter-vacancy response. As a function of vacancy separation, same-sublattice (AA/BB) correlations interfere constructively, leading to oscillatory behavior for zigzag-connected vacancy pairs, whereas for armchair-connected pairs the correlations decay purely exponentially (see Appendix~\ref{app:sublattice}). Both behaviors are direct consequences of the localized nature of the vacancy-induced in-gap states.

By contrast, opposite-sublattice (AB/BA) contributions are strongly suppressed. In this case, destructive interference between different dangling-spin channels leads to substantial cancellation, rendering the net signal small and making its apparent distance dependence particularly sensitive to finite-size effects.

\section{Discussion}
\label{Sec:Conclusion}

In this work, we developed a planar tunneling formalism for a Kitaev spin liquid with dilute spin vacancies and analyzed how fractionalized Majorana excitations manifest in the tunneling response. By decomposing the disorder-averaged spin spectral function into bulk, intra-vacancy, and inter-vacancy components, we identified a clear hierarchy of contributions that governs the in-gap structure and its behavior in the thermodynamic limit.

The bulk sector gives rise to a broad spin-excitation continuum, whereas the intra- and inter-vacancy contributions isolate the localized Majorana modes bound to vacancies. The in-gap and near-zero-bias features of the tunneling spectrum originate from these vacancy-induced modes, with energies and spectral weights that reflect the underlying Majorana fractionalization. Their coherent contribution in the planar geometry enhances their visibility relative to purely local probes.

Taken together, these results demonstrate that Majorana fractionalization in the Kitaev spin liquid can manifest as a coherent and robust signal in a macroscopic junction, offering a realistic route to the experimental identification of spin-liquid fractionalization.

\acknowledgments
We would like  to thank G\'abor B. Hal\'asz for prior collaboration on this topic.
W.L, V.D. and N.B.P. acknowledge the support from NSF DMR-2310318 and  the support of the Minnesota Supercomputing Institute (MSI) at the University of Minnesota. Support for W.-H.K. was provided by the Office of the Vice Chancellor for Research and Graduate Education at the University of Wisconsin–Madison with funding from the Wisconsin Alumni Research Foundation.\\
~~\\

\appendix
\section{Detailed derivation of the planar tunneling formalism}\label{app:C}

In this appendix, we present a step-by-step derivation of the effective
tunneling Hamiltonian and the resulting inelastic current, starting from
the Anderson tunneling Hamiltonian. We perform a
SW transformation to obtain the effective exchange and
potential-scattering amplitudes used in Sec.~\ref{sec:tunneling formalism} of the main text.

\subsection{From Anderson model to effective exchange}\label{APP:A1}

We begin with an Anderson-impurity description of the interface sites ${\bf r}$,
\begin{equation}
\mathcal{H}_{\text{Anderson}} = H_c + H_d + H_g,
\end{equation}
with
\begin{align}
H_c &= \sum_{k,\sigma,\xi}\epsilon_k
\check c_{k\sigma\xi}^\dagger \check c_{k\sigma\xi},\\
H_d &= \sum_{\bf r} \!\left(E_d n_{\bf r} + U n_{{\bf r}\uparrow}n_{{\bf r}\downarrow}\right),\\
H_g &= \sum_{{\bf r},\sigma,\xi}\!\left(g_{\bf r} \check c_{{\bf r}\sigma\xi}^\dagger d_{{\bf r}\sigma}+{\rm H.c.}\right),
\end{align}
and $n_{\bf r}=\sum_\sigma d_{{\bf r}\sigma}^\dagger d_{{\bf r}\sigma}$.
We assume  $E_d<0$ and $E_d+U>0$, 
so that the
singly occupied states $\{|\!\uparrow\rangle,|\!\downarrow\rangle\}$
form a local magnetic moment on each interface site.

To integrate out charge fluctuations we perform a SW
transformation
\begin{equation}
H_{\rm eff}=e^S H e^{-S}=H+[S,H]+\tfrac12[S,[S,H]]+\cdots,
\end{equation}
where the SW generator $S$ satisfies $[H_0,S]=-H_g$ with $H_0=H_c+H_d$.
Its explicit form is
\begin{equation}
S = \sum_{{\bf r},k,\sigma,\xi}
\!\left(
\frac{g_{\bf r} \check c_{k\sigma\xi}^\dagger d_{{\bf r}\sigma}}{E_d-\epsilon_k}
- \frac{g_{\bf r}^{*}d_{{\bf r}\sigma}^\dagger \check c_{k\sigma\xi}}{E_d+U-\epsilon_k}
\right).
\end{equation}
Retaining terms up to $O(g_{\bf r}^2)$ gives
\begin{equation}
H_{\rm eff} = H_c + \tfrac12 [S, H_g].
\end{equation}
Evaluating this commutator produces two types of low-energy processes:
\begin{widetext}
\emph{(i) Elastic scattering}
\begin{equation}
H_t=-\!\!\sum_{{\bf r},{\bf r}',\sigma}\sum_{\xi,\xi'}t_{\bf r}\,
\check c_{{\bf r}\sigma\xi}^\dagger \check c_{{\bf r}'\sigma\xi'},\quad
t_{\bf r}=\frac{|g_{\bf r}|^2}{2}\!\left(\frac{1}{E_d+U}+\frac{1}{E_d}\right),
\end{equation}
which describes elastic potential scattering and contributes only a smooth
background to the tunneling conductance.

\emph{(ii) Inelastic (spin-flip) scattering}
\begin{equation}
= \sum_{{\bf r}}\sum_{\sigma,\sigma'}\sum_{\xi,\xi'}
\mathcal{J}_{\bf r}\,
\check c_{{\bf r}\sigma\xi}^\dagger\,
\boldsymbol{\tau}_{\sigma\sigma'}\,
\check c_{{\bf r}\sigma'\xi'}\!\cdot\!
\boldsymbol{\sigma}_{\bf r},\quad
\mathcal{J}_{\bf r}=\frac{|g_{\bf r}|^2}{2}\!\left(\frac{1}{E_d+U}-\frac{1}{E_d}\right),
\end{equation}
\end{widetext}
which couples the substrate spins to the KSL spins $\boldsymbol\sigma_{\bf r}$.
Collecting the elastic and inelastic terms, the  effective
low-energy tunneling Hamiltonian takes the form
\begin{equation}
\mathcal{H}_{\rm tunnel} = H_t + H_{\mathcal{J}}.
\end{equation}

\subsection{Current operator and linear-response kernel}\label{APP:A2}

We denote by 
$N_1=\sum_{k,\sigma}\check c^\dagger_{k\sigma 1}\check c_{k\sigma 1}$ 
the total electron number in lead~1.  
The current flowing out of that lead follows from charge conservation: it is given
by the rate of change of $N_1$,
\begin{eqnarray}
\hat I(t) = -e\,\dot N_1(t).
\end{eqnarray}
Using the Heisenberg equation of motion, this becomes
\begin{eqnarray}
\hat I(t) = i e\, [\,H_{\rm tunnel},\, N_1\,].
\end{eqnarray}
Only the exchange term $H_{\mathcal J}$ contributes to the inelastic response,
since the potential-scattering term $H_t$ commutes with $N_1$ up to elastic
backscattering.

Because an inelastic process requires one interaction that exchanges energy with the spin system and a second that returns the electron to the lead, the lowest nonvanishing contribution to the inelastic current appears at second order in $\mathcal{J}_r$.
Linear response gives
\begin{eqnarray}
I_{\rm inel}(t)
= i\!\int_{-\infty}^{t}\! dt'\,
\langle [\,\hat I(t), H_{\mathcal{J}}(t')\,]\rangle_0,
\end{eqnarray}
where the average is over the uncoupled leads and Kitaev QSL. 
Using Wick’s theorem for the noninteracting conduction electrons, the
four-fermion correlator generated by the commutator
$[\hat I(t),H_{\mathcal J}(t')]$ factorizes into a product of two
single-particle Green’s functions. In the spin channel this leads to the
standard fermionic particle–hole bubble,
\begin{eqnarray}
\Pi^{R}_{\xi\xi'}(t) &\sim&
{\rm Tr}\!\left[\tau^{\alpha} G_{\xi}(t)\,\tau^{\beta}G_{\xi'}(-t)\right],
\end{eqnarray}
with $G_{\xi}(t)$  denoting the Green’s function in the lead~$\xi$. 
The Fourier transform of this bubble contains an implicit integration
over all momenta in the two–dimensional metal,
\begin{eqnarray}
\Pi^R_{\xi\xi'}(\omega)
=\int\!\frac{d^2q}{(2\pi)^2}\;\Pi^R_{\xi\xi'}(q,\omega),
\end{eqnarray}
which determines the total contribution of particle–hole excitations at
energy~$\omega$.
Carrying out this Fourier transform and keeping only the retarded
component, we obtain
\begin{align}
I_{\rm inel}(V)
&=4\!\sum_{{\bf r},{\bf r}'}\int_{0}^{eV}\!\frac{d\omega}{2\pi}\,
\mathcal{J}_{\bf r}\mathcal{J}_{{\bf r}'}\,
\Im C^R({\bf r},{\bf r}';\omega)\nonumber\\
&\qquad\times
\Big[\Im\Pi^{R}_{12}(eV-\omega)
      -\Im\Pi^{R}_{21}(\omega-eV)\Big],
\label{eq:I_inel}
\end{align}
where
\begin{equation}
C^R(x,x';\omega)
=-i\!\int_0^{\infty}\! dt\, e^{i\omega t}\,
\langle[\hat\sigma_{\bf r}(t),\hat\sigma_{{\bf r}'}(0)]\rangle_{\rm QSL},
\label{eq:CR}
\end{equation}
is the retarded spin correlator of the Kitaev QSL, and
\begin{align}
\Pi^R_{\xi\xi'}({\bf r}-{\bf r}';\omega)
&=-i\!\int_0^{\infty}\! dt\, e^{i\omega t}\nonumber\\
&\times
\langle[
\check{\mathbf c}^\dagger_{{\bf r}\xi}(t)\!\cdot\!\boldsymbol{\tau},
\check{\mathbf c}_{{\bf r}'\xi'}(0)\!\cdot\!\boldsymbol{\tau}
]\rangle_{\rm lead}
\label{eq:PiR}
\end{align}
is the retarded spin bubble of the metallic leads.

For a clean two-dimensional metallic lead with density of states
$\nu_{2D}$ and Fermi velocity $v_F$, the imaginary part of the spin bubble
takes the familiar particle–hole form
\begin{eqnarray}
\Im\Pi^{R}_{12}(q,\omega)
\simeq -\pi \nu_{2D}^2\,\frac{\omega}{v_F q}\,
\theta(v_F q-|\omega|).
\end{eqnarray}
Integrating over momenta in Eq.(\ref{eq:I_inel}) requires an ultraviolet cutoff $q_{\max}$, which
accounts for the finite spatial range of the interface coupling and limits
momentum transfer to values of order the inverse lattice spacing.
With this cutoff,
we obtain
\begin{eqnarray}
\int \frac{d^2 q}{(2\pi)^2}\,
\big|\Im \Pi^{R}_{12}(q,\omega)\big|
\simeq
\left(\frac{\nu_{2D} q_{\max}}{2\pi v_F}\right)^{\!2}
|\omega|.
\end{eqnarray}
Substituting this result into Eq.~(\ref{eq:I_inel}) and differentiating twice
with respect to the bias voltage yields Eq.~(\ref{eq:d2I_main_eqnarray}) of the main text,
\begin{align}
\frac{d^2 I_{\rm inel}}{dV^2}
&= -8
\left(\frac{\nu_{2D} q_{\max}}{2\pi v_F}\right)^{\!2}
\mathcal{J}_0^2 \nonumber\\
&\qquad\times
\sum_{{\bf r},{\bf r}'\notin \mathbb{V}}
\Im C^R({\bf r},{\bf r}';\omega=eV).
\label{eq:d2I}
\end{align}

\section{Spin correlations in the dangling and bulk sectors}
\label{app:B}

% --- Preliminaries and notation ---
%\subsection*{Preliminaries and notation}

This appendix collects the explicit expressions for the spin–spin correlation matrix elements appearing in Sec.~\ref{subsec:vac_S}, which are evaluated by using the Bogoliubov diagonalization of the quadratic Majorana Hamiltonian. We present results separately for dangling (vacancy-induced) and bulk contributions.

We follow Sec.~\ref{sec:spin_correlations} and write the spin operator as
bond fermions. In the clean limit, this implements the bond selection rule $j,k\in\langle jk\rangle_\alpha$.
Moreover, in the planar tunneling geometry only spin-diagonal components $S^{\alpha\alpha}$ enter.
Throughout we pair Majoranas into complex fermions and diagonalize the quadratic Hamiltonian with Bogoliubov matrices \(X,Y\):
\begin{equation}
\begin{pmatrix} f \\ f^\dagger \end{pmatrix}
=
\begin{pmatrix}
X^{\mathsf T} & Y^\dagger\\
Y^{\mathsf T} & X^\dagger
\end{pmatrix}
\begin{pmatrix} a \\ a^\dagger \end{pmatrix}.
%\tag{E.2}
\end{equation}
 We further define $W \equiv X + Y$ and $Z \equiv X - Y$, and denote sites adjacent to vacancies by tildes, e.g., $\tilde b_j^\alpha$ and $\tilde{\sigma}_j^\alpha$.
 In the following, $A$ and $B$ label the two sublattices of the honeycomb lattice. This sublattice classification is independent of whether a site is dangling or non-dangling, which is determined solely by its proximity to a vacancy.

% --- E.1: Dangling sector ---
\subsection{Dangling  spin contributions}

We first consider spin-spin correlations involving only dangling spin components adjacent to vacancies. In this case, the connected
 spin-spin correlation function $S^{\alpha\alpha}_{jk}(\omega)$ can be expressed in the Lehmann representation, where the leading contribution arises from two-particle intermediate states \cite{Kao2024a,Kao2024b}. With $\epsilon_\gamma$ denoting the energy of a one-particle intermediate state $a_{\gamma}$, Eq.~(\ref{eqs:bulk_dangling2}) of the main text can be written explicitly as
\begin{align}
&S^{\alpha\alpha}_{jk}(\omega)
= -\sum_{\gamma,\delta}
\delta\big(\omega-(\epsilon_\gamma+\epsilon_\delta)\big)
\nonumber\\
&\quad\times
\langle\tilde{0}|(\tilde b_j^\alpha c_j)a_\gamma^\dagger a_\delta^\dagger|\tilde{0}\rangle
\langle\tilde{0}|a_\delta a_\gamma(\tilde b_k^\alpha c_k)|\tilde{0}\rangle .
%\tag{E.5}
\label{eqs:dangling_correlation_Lehmann}
\end{align}
These matrix elements can be expressed in closed form in terms of the Bogoliubov matrices $W$ and $Z$. Writing $\ell,m$ ($\ell',m'$) for the $\tilde b^\alpha$ and $c$ sites associated with $j$ ($k$), and denoting
$\delta_{\omega,\gamma\delta}\equiv\delta(\omega-(\epsilon_\gamma+\epsilon_\delta))$,
we get:

\begin{widetext}

\begin{equation}
{\rm{AA:}}\quad S^{\alpha\alpha(2)}_{j\in A,\,k\in A}(\omega)=
2\!\sum_{\gamma,\delta}\!\delta_{\omega,\gamma\delta}\,
\big[ W^{\mathsf T}_{m\gamma}Z^{\mathsf T}_{\ell\delta}\big]\,
\big[ W^\dagger_{m'\gamma}Z^\dagger_{\ell'\delta}-W^\dagger_{m'\delta}Z^\dagger_{\ell'\gamma}\big],
%\tag{E.6}
\label{E6}
\end{equation}
\begin{equation}
 {\rm{BB:}}\quad  S^{\alpha\alpha(2)}_{j\in B,\,k\in B}(\omega)=
2\!\sum_{\gamma,\delta}\!\delta_{\omega,\gamma\delta}\,
\big[ Z^{\mathsf T}_{m\gamma}W^{\mathsf T}_{\ell\delta}\big]\,
\big[ Z^\dagger_{m'\gamma}W^\dagger_{\ell'\delta}-W^\dagger_{\ell'\gamma}Z^\dagger_{m'\delta}\big],
%\tag{E.7}
\label{E7}
\end{equation}
\begin{equation}
{\rm{AB:}}\quad  S^{\alpha\alpha(2)}_{j\in A,\,k\in B}(\omega)=
2\!\sum_{\gamma,\delta}\!\delta_{\omega,\gamma\delta}\,
\big[ W^{\mathsf T}_{m\gamma}Z^{\mathsf T}_{\ell\delta}\big]\,
\big[ Z^\dagger_{m'\gamma}W^\dagger_{\ell'\delta}-W^\dagger_{\ell'\gamma}Z^\dagger_{m'\delta}\big],
%\tag{E.8}
\label{E8}
\end{equation}
\begin{equation}
{\rm{BA:}}\quad S^{\alpha\alpha(2)}_{j\in B,\,k\in A}(\omega)=
2\!\sum_{\gamma,\delta}\!\delta_{\omega,\gamma\delta}\,
\big[ Z^{\mathsf T}_{m\gamma}W^{\mathsf T}_{\ell\delta}\big]\,
\big[ W^\dagger_{m'\gamma}Z^\dagger_{\ell'\delta}-Z^\dagger_{\ell'\gamma}W^\dagger_{m'\delta}\big].
%\tag{E.9}
\label{E9}
\end{equation}
\end{widetext}
The structure of Eqs.~(\ref{E6})–(\ref{E9}) admits a transparent physical interpretation.  For the same-sublattice contributions, AA [Eq.~(\ref{E6})] and BB [Eq.~(\ref{E7})], the indices $j$ and $k$ may correspond either to dangling spin components adjacent to the same vacancy or to sites associated with different vacancies centered on the same sublattice sites. Consequently, these terms contain both intra-vacancy correlations, which are strictly local, and inter-vacancy correlations mediated by the hybridization of vacancy-induced Majorana modes.
In contrast, the mixed-sublattice contributions AB and BA, given by Eqs.~(\ref{E8}) and (\ref{E9}), arise exclusively from inter-vacancy processes. Since a single vacancy generates dangling spin components only on one sublattice, correlations between $A$- and $B$-sublattice dangling spin components necessarily involve two distinct vacancies.

% --- E.2: Bulk sector ---
\subsection{Bulk  spin contributions}

The bulk contributions can be divided into two distinct classes. When both sites $j$ and $k$ are not adjacent to any vacancy, the resulting contribution corresponds to $\overline{S}_{\rm away}$ in Fig.~\ref{fig:flow diagram} and closely resembles the spin correlations of the clean Kitaev model. By contrast, when one of the sites is a  site adjacent to a vacancy and the other is its nearest-neighbor  site, the corresponding contribution is denoted by $\overline{S}_{\rm near}$ in Fig.~\ref{fig:flow diagram}.

In both cases, the key physical process is the same: the spin operator acting on the bulk (i.e., non-dangling) site flips the $\mathbb{Z}_2$ gauge link on the bond $\langle jk\rangle_\alpha$, thereby creating a local two-flux excitation. As a consequence, the spin–spin correlation function reduces to a one-particle Lehmann sum evaluated in the vacancy-induced bound-flux sector. Denoting by $\tilde{\Delta}_{2f}$ the corresponding local two-flux excitation energy and by $\{\epsilon'_\lambda\}$ the single-particle energies in this sector, we obtain
\begin{align}
&S^{\alpha\alpha\,(1)}_{jk}(\omega)
=\xi_{jk}\sum_{\lambda}
\delta\!\big(\omega-(\tilde{\Delta}_{2f}+\epsilon'_\lambda)\big)
\nonumber\\
&\quad\times
\langle \tilde{0}|c_j (a'_\lambda)^\dagger|\tilde{0}\rangle
\langle \tilde{0}|a'_\lambda c_k|\tilde{0}\rangle,
\tag{E.10}
%\label{E10}
\end{align}
with $\xi_{jk}\equiv -\eta_j\eta_k\zeta_j$. 
Expressed in terms of the Bogoliubov matrices $W$ and $Z$, the corresponding one-particle bulk contributions take the form
\begin{align}
&j\in A,k\in A:\;\; \sum_\lambda \delta(\omega-\tilde\Delta_{2f}-\epsilon'_\lambda)\, W^{\mathsf T}_{\ell\lambda}W^\dagger_{m\lambda},
%\tag{E.11}
\label{E11}
\\
&j\in B,k\in B:\;\; \sum_\lambda \delta(\omega-\tilde\Delta_{2f}-\epsilon'_\lambda)\, Z^{\mathsf T}_{\ell\lambda}Z^\dagger_{m\lambda},
%\tag{E.12}
\label{E12}
\\
&j\in A,k\in B:\;\; \sum_\lambda \delta(\omega-\tilde\Delta_{2f}-\epsilon'_\lambda)\, W^{\mathsf T}_{\ell\lambda}Z^\dagger_{m\lambda},
%\tag{E.13}
\label{E13}
\\
&j\in B,k\in A:\;\; \sum_\lambda \delta(\omega-\tilde\Delta_{2f}-\epsilon'_\lambda)\, Z^{\mathsf T}_{\ell\lambda}W^\dagger_{m\lambda}.
%\tag{E.14}
\label{E14}
\end{align}
Here, the same-sublattice contributions, Eqs.~(\ref{E11}) and (\ref{E12}), are nonzero only for $j=k$ and therefore represent purely on-site spin correlations, while the cross-sublattice terms, Eqs.~(\ref{E13}) and (\ref{E14}), are nonzero only for nearest-neighbor pairs.

In the Majorana  formulation, acting with a spin operator on a bulk site flips the $\mathbb{Z}_2$ gauge link on the bond $\langle jk\rangle_\alpha$, or equivalently changes the occupation of the associated bond fermion. For cross-sublattice pairs ($AB/BA$), this operation introduces an additional minus sign in the matrix elements when the bond fermion is occupied, $n_{jk}=1$, due to fermionic anticommutation relations. As a result, only the $AB/BA$ contributions acquire an overall sign change, while the same-sublattice ($AA/BB$) terms remain unaffected, consistent with the general bulk correlation structure discussed in Sec.~\ref{subsec:vac_S}
and summarized in Eq.~(\ref{eq:bulk_correlation_expression}).

\begin{comment}
In the Majorana formulation, flipping the bond $\langle jk\rangle_\alpha$ is accompanied by a change in the occupation of the associated bond fermion, which equivalently tracks the $\mathbb{Z}_2$ gauge link on that bond. For cross-sublattice pairs ($j\in A$, $k\in B$ or vice versa), the product of spin operators involves Majorana fermions residing on the bond $\langle jk\rangle_\alpha$. Reordering these operators into bond and matter fermions requires commuting fermionic operators across the bond. When the corresponding bond fermion is occupied, $n_{jk}=1$, fermionic anticommutation relations produce an additional minus sign in the matrix elements. Consequently, only the cross-sublattice ($AB/BA$) contributions acquire an overall sign change, while same-sublattice ($AA/BB$) terms remain unaffected. This sign structure is consistent with the general bulk correlation expression in Eq.~(\ref{eq:bulk_correlation_expression}).
\end{comment}

% ==================== Appendix C ====================

\section{Scaling and structure of inter-vacancy dangling contributions}
\label{app:sublattice}

\subsection{Scaling of inter-vacancy dangling contributions}
We first analyze the role of inter-vacancy dangling correlations in the dilute-vacancy regime and explain why they remain subdominant compared to intra-vacancy contributions in the thermodynamic limit. To isolate the essential physics, we focus on configurations containing only a small number of vacancies. 

The key physical input is the localized nature of the vacancy-induced in-gap states. Similarly to graphene \cite{Pereira2006,Pereira2008}, when the time-reversal-symmetry-breaking term is vanishingly small, a single vacancy in a honeycomb lattice generates a zero-energy mode whose wave function resides predominantly on the opposite sublattice and decays algebraically away from the vacancy center. In the presence of any finite spectral gap, irrespective of whether it originates from a time-reversal-symmetry-breaking perturbation or from flux excitations, this algebraic decay is cut off at long distances. The vacancy-induced mode then crosses over to an exponential suppression beyond a characteristic length scale~$\xi$. As a result, the envelope of the vacancy-induced wave function scales parametrically as
\begin{equation}
|\Psi(\mathbf{r})| \sim \frac{1}{r}\,e^{-r/\xi},\label{eq:wavefunc_decay}
\end{equation}
where $r$ denotes the distance from the vacancy center.
 This scaling remains valid at finite vacancy densities, as was confirmed numerically in Ref.~\cite{Kao2021vacancy}.

Because inter-vacancy dangling correlations are mediated by the overlap of such localized modes, they inherit the same spatial decay and therefore decrease rapidly with increasing vacancy separation. As a consequence, the contribution from any given vacancy pair remains finite and does not grow with system size. Summing over all vacancy pairs  thus yields a total inter-vacancy contribution that scales at most linearly with the number of sites. 
By contrast, the intra-vacancy contribution is strictly local and additive over vacancies, resulting in a parametrically larger weight in the thermodynamic limit. This hierarchy underlies the dominance of the intra-vacancy term in the dangling contribution discussed in the main text.

\subsection{Spatial structure of inter-vacancy dangling correlations}

\begin{figure*}[t]
  \centering
  \includegraphics[width=1\textwidth]{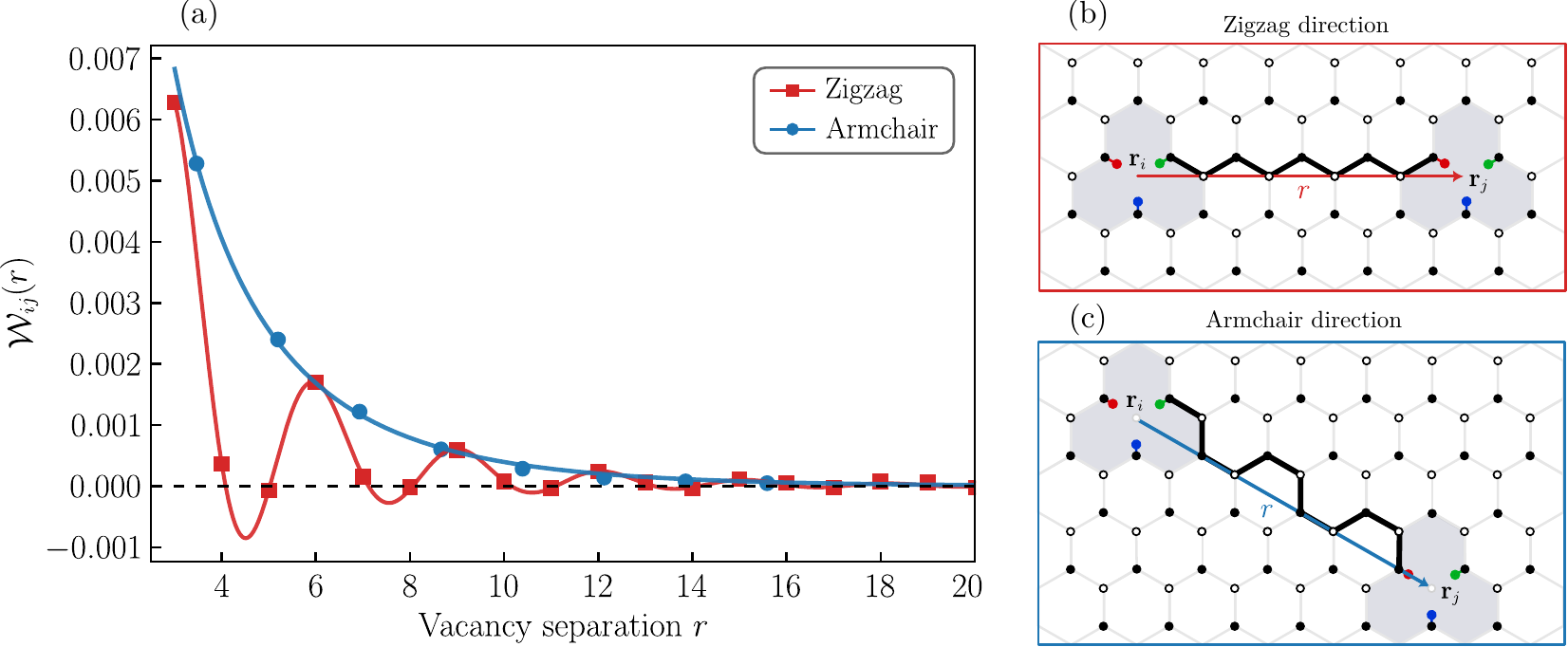}\hfill
  \caption{(a) Sublattice-resolved integrated weight $\mathcal{W}_{ij}(r)$ for same-sublattice (AA/BB) versus distance $r = |\br_i - \br_j|$ (in unit cell units). The zigzag direction (filled squares) shows a period-three
  modulation atop an exponential envelope. In this case the fitting parameters in Eq. (\ref{eq:fitAA_app}) are $s = 1.0$, $A = 0.02$, $\xi = 4.92$ and $C = 0.56$. For the armchair case (filled circles) the fitting parameters are $s=1.0$, $A = 0.37$, $\xi = 4.26$ and $C = 0.11$. 
  Zigzag (b) and Armchair (c) directions of vacancy pairs at $\bm{r}_i,\bm{r}_j$ are presented on the right panel. } 
  \label{fig:appC_osc}
\end{figure*}

To  characterize the contributions from the near-zero-energy vacancy-induced modes into the inter-vacancy dangling correlations, we introduce the low-energy spectral weight
\begin{equation}
\mathcal{W}_{ij}^{\alpha\alpha}
=
\int_{0}^{\omega_0}\! d\omega\;
S^{\alpha\alpha}_{\bm{r}_i+\bm{\delta}_{\alpha},\,\bm{r}_j+\bm{\delta}_{\alpha}}(\omega),
\label{eq:Inz_def}
\end{equation}
which measures the integrated subgap contribution of the dangling-spin correlation function between sites associated with vacancies at positions $\bm{r}_i$ and $\bm{r}_j$, as shown for a vacancy pair in Fig. \ref{fig:appC_osc}(b) and (c) for zigzag and armchair directions, respectively. The upper cutoff $\omega_0/J = 0.05$
is chosen to isolate the contribution of low-energy in-gap states (see Fig.~\ref{fig:wide}). The bond-direction vector $\bm{\delta}_{\alpha}$ ($\alpha=x,y,z$) is included to account for the displacement of the dangling Majorana mode relative to the vacancy center. 

In Fig.~\ref{fig:appC_osc} (a),  we plot $\mathcal{W}_{ij}(r)\equiv\sum_\alpha\mathcal{W}_{ij}^{\alpha\alpha}$  as a function of the vacancy separation  $r = |\bm{r}_i - \bm{r}_j|$ for same-sublattice vacancy pairs (AA/BB) connected along zigzag and armchair directions. For zigzag-connected pairs, the spectral weight displays pronounced oscillations with distance, superimposed on an overall exponential decay. In contrast, for armchair-connected pairs, it decays monotonically without visible oscillations. The opposite-sublattice contribution (not shown) is suppressed by roughly an order of magnitude and alternates in sign with separation, leading to strong destructive interference. Consequently, we focus only on same-sublattice dangling correlations.

To qualitatively understand the distance dependence of the inter-vacancy dangling spin correlation function, we consider $S^{\al\al}_{ij}(\om)$ in the Lehmann representation with two-particle intermediate states shown in Eq.~(\ref{eqs:dangling_correlation_Lehmann}).  To have a finite correlation below the bulk energy gap, we consider that the wavefunction associated to $a^{\dg}_{\gm}$ is localized around site $i$, with a tiny overlap with the Majorana wavefunctions on site $j$:
\begin{align}
\begin{split}
&a_{\gm}^{\dg} \sim \eta^{\phantom{\dg}}_{1} \tilde{b}_i^{\al} + \eta^{\phantom{\dg}}_{2}c^{\phantom{\dg}}_i + \chi^{\phantom{\dg}}_1\tilde{b}_{j}^{\al}+\chi^{\phantom{\dg}}_2c^{\phantom{\dg}}_j\\
\end{split}
\end{align}
where the coefficients $\eta \gg \chi$ due to its localized nature. Therefore, the remaining terms in the four-particle expectation values are
\begin{align}
\begin{split}
&\big\langle\tilde{b}_{i}^{\al}c^{\phantom{\dg}}_i a_{\gm}^{\dg}a_{\dt}^{\dg}\big\rangle \sim -\eta^{\phantom{\dg}}_1\big\langle c^{\phantom{\dg}}_i a_{\dt}^{\dg}\big\rangle+\eta^{\phantom{\dg}}_{2}\big\langle\tilde{b}_i^{\al}a_{\dt}^{\dg}\big\rangle\\
&\big\langle a^{\phantom{\dg}}_{\dt}a^{\phantom{\dg}}_{\gm}\tilde{b}_{j}^{\al}c^{\phantom{\dg}}_j\big\rangle\sim \chi^{*}_1 \big\langle a^{\phantom{\dg}}_{\dt}c^{\phantom{\dg}}_j \big\rangle - \chi^{*}_2 \big\langle a^{\phantom{\dg}}_{\dt}\tilde{b}^{\al}_j\big\rangle,
\end{split}
\end{align}
Since we treat the dangling Majorana fermions as additional itinerant particles in the system, the product of the these correlators can be written as
\begin{align}
\big\langle\tilde{b}_{i}^{\al}c^{\phantom{\dg}}_i a_{\gm}^{\dg}a_{\dt}^{\dg}\big\rangle\big\langle a^{\phantom{\dg}}_{\dt}a^{\phantom{\dg}}_{\gm}\tilde{b}_{j}^{\al}c^{\phantom{\dg}}_j\big\rangle\sim \sum_{\mu,\nu}\eta^{\phantom{\dg}}_{\mu}\chi^{*}_{\nu}\big\langle d_i^{\mu}a_{\dt}^{\dg}\big\rangle \big\langle a^{\phantom{\dg}}_{\dt}d_j^{\nu}\big\rangle,
\end{align}
where $d^{\mu}_i$ is a generalized Majorana operator such that $d^{1}_i = c^{\phantom{\dg}}_i$  and $d_i^{2} = \tilde{b}_i^{\al}$. The dangling correlation function becomes
\begin{align}
S^{\al\al}_{ij}(\om') \sim -\sum_{\dt}  \sum_{\mu,\nu}\eta^{\phantom{\dg}}_{\mu}\chi^{*}_{\nu}\big\langle d_i^{\mu}a_{\dt}^{\dg}\big\rangle\big\langle a^{\phantom{\dg}}_{\dt}d_j^{\nu}\big\rangle\,\dt(\om'-\ep_{\dt}),
\end{align}
where $\om'$ has a tiny shift from $\om$ by the specific $\ep_{\gm}$ of the localized mode in the low-energy manifold. The general expression of this inter-vacancy correlation function is the local two-point correlator between site $i$ and $j$. Considering that the low-energy description of our model is a massive Dirac Hamiltonian with two valleys, one can show that in the continuum limit
\begin{align}
\big\langle d^{\mu}_i a_{\dt}^{\dg}\big\rangle \sim e^{i\bK\cdot \br_i} F^{\mu}_{\bK, \dt}(\br_i) + e^{i\bK'\cdot \br_i} F^{\mu}_{\bK',\dt}(\br_i),  
\end{align}
where $\bK$ and $\bK'$ are the two valleys and $F_{\bK}$ and $F_{\bK'}$ are the envelope functions. Therefore, the product $\big\langle d_i^{\mu}a_{\dt}^{\dg}\big\rangle\big\langle a^{\phantom{\dg}}_{\dt}d_j^{\nu}\big\rangle$ contains the intra-valley part as $\sim \cos(\bK\cdot \br)$ and the inter-valley part as $\sim \cos(\bK\cdot \bR+\phi_{\dt})$, where $\br = \br_{i}-\br_{j}$ is the distance of separation and $\bR = \br_i+\br_j$. Note that the phase shift $\phi_{\dt}$ of the inter-valley oscillation is frequency-dependent and is averaged out in the integrated spectral weight $\mathcal{W}_{ij}^{\al\al}(r)$. Since $\br$ is a Bravais lattice vector and $\bK = \frac{4\pi}{3a}(1,0)$, we know that
\begin{align}
\cos(\bK\cdot \br ) = \cos(-\bK\cdot\br) = \cos(2\bK\cdot \br),
\end{align}
which captures the oscillatory part of the integrated weight. In addition, the envelope function provides the exponential decay in $\br$ due to the localized nature of the in-gap states.

As a result, the spatial dependence of the inter-vacancy response reflects an interplay between exponential suppression set by the bulk gap and oscillatory behavior inherited from the valley structure of the low-energy Majorana modes. This motivates us to fit the numerically obtained distance dependence of $\mathcal{W}_{ij}(r)$ using a simple valley-interference envelope of the form \cite{Saremi2007_RKKY_Graphene, Sherafati2011_RKKY, Kogan2013_RKKY_Gapped_Doped_Graphene}
\begin{equation}
\mathcal{W}_{ij}(r)= A\, r^{-s} e^{-r/\xi}
\big[C+\cos\!\big(2\bK\!\cdot\!\br\big)\big],
\label{eq:fitAA_app}
\end{equation}
where $s$, $A$, $C$ and $\xi$ are fitting parameters.
%The origin of the oscillatory term in Eq.~(\ref{eq:fitAA_app}) can be understood as follows. Interference between contributions associated with the two inequivalent valleys naturally produces terms proportional to $\cos(2\bm K\!\cdot\!\bm r)$. 
Notably, whether the oscillatory part survives in the integrated
spectral weight $\mathcal{W}^{\alpha\alpha}_{ij}$ depends sensitively on the
geometry of the vacancy separation. For the zigzag-connected vacancy pairs along the
direction $\bm a_1=a(1,0)$
[Fig.~\ref{fig:appC_osc} (b)], the oscillatory factor becomes
\begin{equation*}
\cos\!\big(2\bK\!\cdot\!\br\big)
=
\cos\!\left(\tfrac{4\pi}{3}n\right), \qquad n\in\mathbb Z ,
\end{equation*}
leading to pronounced oscillations superimposed on the overall exponential decay
set by the localized envelope of the vacancy-induced modes. By contrast, for armchair-connected vacancy pairs the valley phase factors cancel.
Along the direction $2\bm a_1-\bm a_2=a(\tfrac{3}{2},\tfrac{\sqrt{3}}{2})$, one finds
\begin{equation*}
\cos\!\big(2\bK\!\cdot\!\br\big)
=
\cos\!\left(4\pi n\right)=1,
\end{equation*}
so that the oscillatory contribution is eliminated, leaving a purely exponential
decay governed by the localized envelope of the vacancy-induced modes.

In this way, the geometry of the vacancy separation determines whether or not the oscillatory term contributes to $\mathcal{W}^{\alpha\alpha}_{ij}(r)$: zigzag-connected vacancy pairs exhibit pronounced oscillations, while for armchair-connected pairs the interference reduces to a constant and the spatial dependence is controlled entirely by the non-oscillatory $r^{-s}e^{-r/\xi}$ envelope.

\bibliography{references}% Produces the bibliography via BibTeX.

%apsrev4-2.bst 2019-01-14 (MD) hand-edited version of apsrev4-1.bst
%Control: key (0)
%Control: author (8) initials jnrlst
%Control: editor formatted (1) identically to author
%Control: production of article title (0) allowed
%Control: page (0) single
%Control: year (1) truncated
%Control: production of eprint (0) enabled
\begin{thebibliography}{53}%
\makeatletter
\providecommand \@ifxundefined [1]{%
 \@ifx{#1\undefined}
}%
\providecommand \@ifnum [1]{%
 \ifnum #1\expandafter \@firstoftwo
 \else \expandafter \@secondoftwo
 \fi
}%
\providecommand \@ifx [1]{%
 \ifx #1\expandafter \@firstoftwo
 \else \expandafter \@secondoftwo
 \fi
}%
\providecommand \natexlab [1]{#1}%
\providecommand \enquote  [1]{``#1''}%
\providecommand \bibnamefont  [1]{#1}%
\providecommand \bibfnamefont [1]{#1}%
\providecommand \citenamefont [1]{#1}%
\providecommand \href@noop [0]{\@secondoftwo}%
\providecommand \href [0]{\begingroup \@sanitize@url \@href}%
\providecommand \@href[1]{\@@startlink{#1}\@@href}%
\providecommand \@@href[1]{\endgroup#1\@@endlink}%
\providecommand \@sanitize@url [0]{\catcode `\\12\catcode `\$12\catcode
  `\&12\catcode `\#12\catcode `\^12\catcode `\_12\catcode `\%12\relax}%
\providecommand \@@startlink[1]{}%
\providecommand \@@endlink[0]{}%
\providecommand \url  [0]{\begingroup\@sanitize@url \@url }%
\providecommand \@url [1]{\endgroup\@href {#1}{\urlprefix }}%
\providecommand \urlprefix  [0]{URL }%
\providecommand \Eprint [0]{\href }%
\providecommand \doibase [0]{https://doi.org/}%
\providecommand \selectlanguage [0]{\@gobble}%
\providecommand \bibinfo  [0]{\@secondoftwo}%
\providecommand \bibfield  [0]{\@secondoftwo}%
\providecommand \translation [1]{[#1]}%
\providecommand \BibitemOpen [0]{}%
\providecommand \bibitemStop [0]{}%
\providecommand \bibitemNoStop [0]{.\EOS\space}%
\providecommand \EOS [0]{\spacefactor3000\relax}%
\providecommand \BibitemShut  [1]{\csname bibitem#1\endcsname}%
\let\auto@bib@innerbib\@empty
%</preamble>
\bibitem [{\citenamefont {Anderson}(1973)}]{anderson1973resonating}%
  \BibitemOpen
  \bibfield  {author} {\bibinfo {author} {\bibfnamefont {P.~W.}\ \bibnamefont
  {Anderson}},\ }\bibfield  {title} {\bibinfo {title} {{Resonating valence
  bonds: A new kind of insulator?}},\ }\href
  {https://www.sciencedirect.com/science/article/pii/0025540873901670}
  {\bibfield  {journal} {\bibinfo  {journal} {Mater. Res. Bull.}\ }\textbf
  {\bibinfo {volume} {8}},\ \bibinfo {pages} {153} (\bibinfo {year}
  {1973})}\BibitemShut {NoStop}%
\bibitem [{\citenamefont {Kitaev}(2006)}]{Kitaev2006}%
  \BibitemOpen
  \bibfield  {author} {\bibinfo {author} {\bibfnamefont {A.}~\bibnamefont
  {Kitaev}},\ }\bibfield  {title} {\bibinfo {title} {{Anyons in an exactly
  solved model and beyond}},\ }\href
  {https://doi.org/10.1016/j.aop.2005.10.005} {\bibfield  {journal} {\bibinfo
  {journal} {Annals of Physics}\ }\textbf {\bibinfo {volume} {321}},\ \bibinfo
  {pages} {2} (\bibinfo {year} {2006})}\BibitemShut {NoStop}%
\bibitem [{\citenamefont {Balents}(2010)}]{Balents2010}%
  \BibitemOpen
  \bibfield  {author} {\bibinfo {author} {\bibfnamefont {L.}~\bibnamefont
  {Balents}},\ }\bibfield  {title} {\bibinfo {title} {{Spin liquids in
  frustrated magnets}},\ }\href@noop {} {\bibfield  {journal} {\bibinfo
  {journal} {Nature}\ }\textbf {\bibinfo {volume} {464}},\ \bibinfo {pages}
  {199} (\bibinfo {year} {2010})}\BibitemShut {NoStop}%
\bibitem [{\citenamefont {Savary}\ and\ \citenamefont
  {Balents}(2016)}]{Savary2016}%
  \BibitemOpen
  \bibfield  {author} {\bibinfo {author} {\bibfnamefont {L.}~\bibnamefont
  {Savary}}\ and\ \bibinfo {author} {\bibfnamefont {L.}~\bibnamefont
  {Balents}},\ }\bibfield  {title} {\bibinfo {title} {{Quantum spin liquids: a
  review}},\ }\href {https://doi.org/10.1088/0034-4885/80/1/016502} {\bibfield
  {journal} {\bibinfo  {journal} {Reports on Progress in Physics}\ }\textbf
  {\bibinfo {volume} {80}},\ \bibinfo {pages} {016502} (\bibinfo {year}
  {2016})}\BibitemShut {NoStop}%
\bibitem [{\citenamefont {Knolle}\ and\ \citenamefont
  {Moessner}(2019)}]{KnolleMoessner2019}%
  \BibitemOpen
  \bibfield  {author} {\bibinfo {author} {\bibfnamefont {J.}~\bibnamefont
  {Knolle}}\ and\ \bibinfo {author} {\bibfnamefont {R.}~\bibnamefont
  {Moessner}},\ }\bibfield  {title} {\bibinfo {title} {{A Field Guide to Spin
  Liquids}},\ }\href {https://doi.org/10.1146/annurev-conmatphys-031218-013401}
  {\bibfield  {journal} {\bibinfo  {journal} {Annu. Rev. Condens. Matter
  Phys.}\ }\textbf {\bibinfo {volume} {10}},\ \bibinfo {pages} {451} (\bibinfo
  {year} {2019})}\BibitemShut {NoStop}%
\bibitem [{\citenamefont {Takagi}\ \emph {et~al.}(2019)\citenamefont {Takagi},
  \citenamefont {Takayama}, \citenamefont {Jackeli}, \citenamefont
  {Khaliullin},\ and\ \citenamefont {Nagaosa}}]{Takagi2019}%
  \BibitemOpen
  \bibfield  {author} {\bibinfo {author} {\bibfnamefont {H.}~\bibnamefont
  {Takagi}}, \bibinfo {author} {\bibfnamefont {T.}~\bibnamefont {Takayama}},
  \bibinfo {author} {\bibfnamefont {G.}~\bibnamefont {Jackeli}}, \bibinfo
  {author} {\bibfnamefont {G.}~\bibnamefont {Khaliullin}},\ and\ \bibinfo
  {author} {\bibfnamefont {N.}~\bibnamefont {Nagaosa}},\ }\bibfield  {title}
  {\bibinfo {title} {{Concept and realization of Kitaev quantum spin
  liquids}},\ }\href {https://doi.org/10.1038/s42254-019-0058-3} {\bibfield
  {journal} {\bibinfo  {journal} {Nature Reviews Physics}\ }\textbf {\bibinfo
  {volume} {1}},\ \bibinfo {pages} {264} (\bibinfo {year} {2019})}\BibitemShut
  {NoStop}%
\bibitem [{\citenamefont {Broholm}\ \emph {et~al.}(2020)\citenamefont
  {Broholm}, \citenamefont {Cava}, \citenamefont {Kivelson}, \citenamefont
  {Nocera}, \citenamefont {Norman},\ and\ \citenamefont
  {Senthil}}]{Broholm2020}%
  \BibitemOpen
  \bibfield  {author} {\bibinfo {author} {\bibfnamefont {C.}~\bibnamefont
  {Broholm}}, \bibinfo {author} {\bibfnamefont {R.~J.}\ \bibnamefont {Cava}},
  \bibinfo {author} {\bibfnamefont {S.~A.}\ \bibnamefont {Kivelson}}, \bibinfo
  {author} {\bibfnamefont {D.~G.}\ \bibnamefont {Nocera}}, \bibinfo {author}
  {\bibfnamefont {M.~R.}\ \bibnamefont {Norman}},\ and\ \bibinfo {author}
  {\bibfnamefont {T.}~\bibnamefont {Senthil}},\ }\bibfield  {title} {\bibinfo
  {title} {{Quantum spin liquids}},\ }\href
  {https://science.sciencemag.org/content/367/6475/eaay0668} {\bibfield
  {journal} {\bibinfo  {journal} {Science}\ }\textbf {\bibinfo {volume} {367}}
  (\bibinfo {year} {2020})}\BibitemShut {NoStop}%
\bibitem [{\citenamefont {Kogut}(1979)}]{Kogut1979}%
  \BibitemOpen
  \bibfield  {author} {\bibinfo {author} {\bibfnamefont {J.~B.}\ \bibnamefont
  {Kogut}},\ }\bibfield  {title} {\bibinfo {title} {{An introduction to lattice
  gauge theory and spin systems}},\ }\href
  {https://doi.org/10.1103/RevModPhys.51.659} {\bibfield  {journal} {\bibinfo
  {journal} {Rev. Mod. Phys.}\ }\textbf {\bibinfo {volume} {51}},\ \bibinfo
  {pages} {659} (\bibinfo {year} {1979})}\BibitemShut {NoStop}%
\bibitem [{\citenamefont {Winter}\ \emph {et~al.}(2017)\citenamefont {Winter},
  \citenamefont {Tsirlin}, \citenamefont {Daghofer}, \citenamefont {van~den
  Brink}, \citenamefont {Singh}, \citenamefont {Gegenwart},\ and\ \citenamefont
  {Valentí}}]{Winter2017}%
  \BibitemOpen
  \bibfield  {author} {\bibinfo {author} {\bibfnamefont {S.~M.}\ \bibnamefont
  {Winter}}, \bibinfo {author} {\bibfnamefont {A.~A.}\ \bibnamefont {Tsirlin}},
  \bibinfo {author} {\bibfnamefont {M.}~\bibnamefont {Daghofer}}, \bibinfo
  {author} {\bibfnamefont {J.}~\bibnamefont {van~den Brink}}, \bibinfo {author}
  {\bibfnamefont {Y.}~\bibnamefont {Singh}}, \bibinfo {author} {\bibfnamefont
  {P.}~\bibnamefont {Gegenwart}},\ and\ \bibinfo {author} {\bibfnamefont
  {R.}~\bibnamefont {Valentí}},\ }\bibfield  {title} {\bibinfo {title}
  {{Models and materials for generalized Kitaev magnetism}},\ }\href
  {https://doi.org/10.1088/1361-648X/aa8cf5} {\bibfield  {journal} {\bibinfo
  {journal} {Journal of Physics: Condensed Matter}\ }\textbf {\bibinfo {volume}
  {29}},\ \bibinfo {pages} {493002} (\bibinfo {year} {2017})}\BibitemShut
  {NoStop}%
\bibitem [{\citenamefont {Hermanns}\ \emph {et~al.}(2018)\citenamefont
  {Hermanns}, \citenamefont {Kimchi},\ and\ \citenamefont
  {Knolle}}]{Hermanns2018}%
  \BibitemOpen
  \bibfield  {author} {\bibinfo {author} {\bibfnamefont {M.}~\bibnamefont
  {Hermanns}}, \bibinfo {author} {\bibfnamefont {I.}~\bibnamefont {Kimchi}},\
  and\ \bibinfo {author} {\bibfnamefont {J.}~\bibnamefont {Knolle}},\
  }\bibfield  {title} {\bibinfo {title} {{Physics of the Kitaev Model:
  Fractionalization, Dynamical Correlations, and Material Connections}},\
  }\href {https://doi.org/10.1146/annurev-conmatphys-033117-053934} {\bibfield
  {journal} {\bibinfo  {journal} {Annu. Rev. Condens. Matter Phys.}\ }\textbf
  {\bibinfo {volume} {9}},\ \bibinfo {pages} {17} (\bibinfo {year}
  {2018})}\BibitemShut {NoStop}%
\bibitem [{\citenamefont {Motome}\ and\ \citenamefont
  {Nasu}(2020)}]{Motome2019}%
  \BibitemOpen
  \bibfield  {author} {\bibinfo {author} {\bibfnamefont {Y.}~\bibnamefont
  {Motome}}\ and\ \bibinfo {author} {\bibfnamefont {J.}~\bibnamefont {Nasu}},\
  }\bibfield  {title} {\bibinfo {title} {{Hunting Majorana Fermions in Kitaev
  Magnets}},\ }\href {https://doi.org/10.7566/JPSJ.89.012002} {\bibfield
  {journal} {\bibinfo  {journal} {J. Phys. Soc. Jpn}\ }\textbf {\bibinfo
  {volume} {89}},\ \bibinfo {pages} {012002} (\bibinfo {year}
  {2020})}\BibitemShut {NoStop}%
\bibitem [{\citenamefont {Trebst}\ and\ \citenamefont
  {Hickey}(2022)}]{Trebst2022}%
  \BibitemOpen
  \bibfield  {author} {\bibinfo {author} {\bibfnamefont {S.}~\bibnamefont
  {Trebst}}\ and\ \bibinfo {author} {\bibfnamefont {C.}~\bibnamefont
  {Hickey}},\ }\bibfield  {title} {\bibinfo {title} {{Kitaev materials}},\
  }\href {https://doi.org/https://doi.org/10.1016/j.physrep.2021.11.003}
  {\bibfield  {journal} {\bibinfo  {journal} {Physics Reports}\ }\textbf
  {\bibinfo {volume} {950}},\ \bibinfo {pages} {1} (\bibinfo {year}
  {2022})}\BibitemShut {NoStop}%
\bibitem [{\citenamefont {Nasu}(2023)}]{Nasu2023}%
  \BibitemOpen
  \bibfield  {author} {\bibinfo {author} {\bibfnamefont {J.}~\bibnamefont
  {Nasu}},\ }\bibfield  {title} {\bibinfo {title} {{Majorana quasiparticles
  emergent in Kitaev spin liquid}},\ }\href
  {https://doi.org/10.1093/ptep/ptad115} {\bibfield  {journal} {\bibinfo
  {journal} {Prog. Theor. Exp. Phys.}\ }\textbf {\bibinfo {volume} {2024}},\
  \bibinfo {pages} {08C104} (\bibinfo {year} {2023})}\BibitemShut {NoStop}%
\bibitem [{\citenamefont {Matsuda}\ \emph {et~al.}(2025)\citenamefont
  {Matsuda}, \citenamefont {Shibauchi},\ and\ \citenamefont
  {Kee}}]{MatsudaRMP2025}%
  \BibitemOpen
  \bibfield  {author} {\bibinfo {author} {\bibfnamefont {Y.}~\bibnamefont
  {Matsuda}}, \bibinfo {author} {\bibfnamefont {T.}~\bibnamefont {Shibauchi}},\
  and\ \bibinfo {author} {\bibfnamefont {H.-Y.}\ \bibnamefont {Kee}},\
  }\bibfield  {title} {\bibinfo {title} {{Kitaev quantum spin liquids}},\
  }\href {https://doi.org/10.1103/3m4m-3v59} {\bibfield  {journal} {\bibinfo
  {journal} {Rev. Mod. Phys.}\ }\textbf {\bibinfo {volume} {97}},\ \bibinfo
  {pages} {045003} (\bibinfo {year} {2025})}\BibitemShut {NoStop}%
\bibitem [{\citenamefont {Willans}\ \emph {et~al.}(2010)\citenamefont
  {Willans}, \citenamefont {Chalker},\ and\ \citenamefont
  {Moessner}}]{Willans2010}%
  \BibitemOpen
  \bibfield  {author} {\bibinfo {author} {\bibfnamefont {A.~J.}\ \bibnamefont
  {Willans}}, \bibinfo {author} {\bibfnamefont {J.~T.}\ \bibnamefont
  {Chalker}},\ and\ \bibinfo {author} {\bibfnamefont {R.}~\bibnamefont
  {Moessner}},\ }\bibfield  {title} {\bibinfo {title} {{Disorder in a Quantum
  Spin Liquid: Flux Binding and Local Moment Formation}},\ }\href
  {https://doi.org/10.1103/PhysRevLett.104.237203} {\bibfield  {journal}
  {\bibinfo  {journal} {Phys. Rev. Lett.}\ }\textbf {\bibinfo {volume} {104}},\
  \bibinfo {pages} {237203} (\bibinfo {year} {2010})}\BibitemShut {NoStop}%
\bibitem [{\citenamefont {Willans}\ \emph {et~al.}(2011)\citenamefont
  {Willans}, \citenamefont {Chalker},\ and\ \citenamefont
  {Moessner}}]{Willans2011}%
  \BibitemOpen
  \bibfield  {author} {\bibinfo {author} {\bibfnamefont {A.~J.}\ \bibnamefont
  {Willans}}, \bibinfo {author} {\bibfnamefont {J.~T.}\ \bibnamefont
  {Chalker}},\ and\ \bibinfo {author} {\bibfnamefont {R.}~\bibnamefont
  {Moessner}},\ }\bibfield  {title} {\bibinfo {title} {{Site dilution in the
  Kitaev honeycomb model}},\ }\href
  {https://doi.org/10.1103/PhysRevB.84.115146} {\bibfield  {journal} {\bibinfo
  {journal} {Phys. Rev. B}\ }\textbf {\bibinfo {volume} {84}},\ \bibinfo
  {pages} {115146} (\bibinfo {year} {2011})}\BibitemShut {NoStop}%
\bibitem [{\citenamefont {Knolle}\ \emph {et~al.}(2019)\citenamefont {Knolle},
  \citenamefont {Moessner},\ and\ \citenamefont {Perkins}}]{Knolle2019}%
  \BibitemOpen
  \bibfield  {author} {\bibinfo {author} {\bibfnamefont {J.}~\bibnamefont
  {Knolle}}, \bibinfo {author} {\bibfnamefont {R.}~\bibnamefont {Moessner}},\
  and\ \bibinfo {author} {\bibfnamefont {N.~B.}\ \bibnamefont {Perkins}},\
  }\bibfield  {title} {\bibinfo {title} {{Bond-Disordered Spin Liquid and the
  Honeycomb Iridate ${\mathrm{H}}_{3}{\mathrm{LiIr}}_{2}{\mathrm{O}}_{6}$:
  Abundant Low-Energy Density of States from Random Majorana Hopping}},\ }\href
  {https://doi.org/10.1103/PhysRevLett.122.047202} {\bibfield  {journal}
  {\bibinfo  {journal} {Phys. Rev. Lett.}\ }\textbf {\bibinfo {volume} {122}},\
  \bibinfo {pages} {047202} (\bibinfo {year} {2019})}\BibitemShut {NoStop}%
\bibitem [{\citenamefont {Nasu}\ and\ \citenamefont {Motome}(2020)}]{Nasu2020}%
  \BibitemOpen
  \bibfield  {author} {\bibinfo {author} {\bibfnamefont {J.}~\bibnamefont
  {Nasu}}\ and\ \bibinfo {author} {\bibfnamefont {Y.}~\bibnamefont {Motome}},\
  }\bibfield  {title} {\bibinfo {title} {{Thermodynamic and transport
  properties in disordered Kitaev models}},\ }\href
  {https://doi.org/10.1103/PhysRevB.102.054437} {\bibfield  {journal} {\bibinfo
   {journal} {Phys. Rev. B}\ }\textbf {\bibinfo {volume} {102}},\ \bibinfo
  {pages} {054437} (\bibinfo {year} {2020})}\BibitemShut {NoStop}%
\bibitem [{\citenamefont {Nasu}\ and\ \citenamefont {Motome}(2021)}]{Nasu2021}%
  \BibitemOpen
  \bibfield  {author} {\bibinfo {author} {\bibfnamefont {J.}~\bibnamefont
  {Nasu}}\ and\ \bibinfo {author} {\bibfnamefont {Y.}~\bibnamefont {Motome}},\
  }\bibfield  {title} {\bibinfo {title} {{Spin dynamics in the Kitaev model
  with disorder: Quantum Monte Carlo study of dynamical spin structure factor,
  magnetic susceptibility, and NMR relaxation rate}},\ }\href
  {https://doi.org/10.1103/PhysRevB.104.035116} {\bibfield  {journal} {\bibinfo
   {journal} {Phys. Rev. B}\ }\textbf {\bibinfo {volume} {104}},\ \bibinfo
  {pages} {035116} (\bibinfo {year} {2021})}\BibitemShut {NoStop}%
\bibitem [{\citenamefont {Kao}\ \emph {et~al.}(2021)\citenamefont {Kao},
  \citenamefont {Knolle}, \citenamefont {Hal\'asz}, \citenamefont {Moessner},\
  and\ \citenamefont {Perkins}}]{Kao2021vacancy}%
  \BibitemOpen
  \bibfield  {author} {\bibinfo {author} {\bibfnamefont {W.-H.}\ \bibnamefont
  {Kao}}, \bibinfo {author} {\bibfnamefont {J.}~\bibnamefont {Knolle}},
  \bibinfo {author} {\bibfnamefont {G.~B.}\ \bibnamefont {Hal\'asz}}, \bibinfo
  {author} {\bibfnamefont {R.}~\bibnamefont {Moessner}},\ and\ \bibinfo
  {author} {\bibfnamefont {N.~B.}\ \bibnamefont {Perkins}},\ }\bibfield
  {title} {\bibinfo {title} {{Vacancy-Induced Low-Energy Density of States in
  the Kitaev Spin Liquid}},\ }\href
  {https://doi.org/10.1103/PhysRevX.11.011034} {\bibfield  {journal} {\bibinfo
  {journal} {Phys. Rev. X}\ }\textbf {\bibinfo {volume} {11}},\ \bibinfo
  {pages} {011034} (\bibinfo {year} {2021})}\BibitemShut {NoStop}%
\bibitem [{\citenamefont {Kao}\ and\ \citenamefont
  {Perkins}(2021)}]{Kao2021localization}%
  \BibitemOpen
  \bibfield  {author} {\bibinfo {author} {\bibfnamefont {W.-H.}\ \bibnamefont
  {Kao}}\ and\ \bibinfo {author} {\bibfnamefont {N.~B.}\ \bibnamefont
  {Perkins}},\ }\bibfield  {title} {\bibinfo {title} {{Disorder upon disorder:
  Localization effects in the Kitaev spin liquid}},\ }\href
  {https://doi.org/https://doi.org/10.1016/j.aop.2021.168506} {\bibfield
  {journal} {\bibinfo  {journal} {Ann. Phys.}\ }\textbf {\bibinfo {volume}
  {435}},\ \bibinfo {pages} {168506} (\bibinfo {year} {2021})}\BibitemShut
  {NoStop}%
\bibitem [{\citenamefont {Dantas}\ and\ \citenamefont
  {Andrade}(2022)}]{Dantas2022}%
  \BibitemOpen
  \bibfield  {author} {\bibinfo {author} {\bibfnamefont {V.}~\bibnamefont
  {Dantas}}\ and\ \bibinfo {author} {\bibfnamefont {E.~C.}\ \bibnamefont
  {Andrade}},\ }\bibfield  {title} {\bibinfo {title} {{Disorder, Low-Energy
  Excitations, and Topology in the Kitaev Spin Liquid}},\ }\href
  {https://doi.org/10.1103/PhysRevLett.129.037204} {\bibfield  {journal}
  {\bibinfo  {journal} {Phys. Rev. Lett.}\ }\textbf {\bibinfo {volume} {129}},\
  \bibinfo {pages} {037204} (\bibinfo {year} {2022})}\BibitemShut {NoStop}%
\bibitem [{\citenamefont {Kao}\ \emph {et~al.}(2024{\natexlab{a}})\citenamefont
  {Kao}, \citenamefont {Perkins},\ and\ \citenamefont {Hal\'asz}}]{Kao2024a}%
  \BibitemOpen
  \bibfield  {author} {\bibinfo {author} {\bibfnamefont {W.-H.}\ \bibnamefont
  {Kao}}, \bibinfo {author} {\bibfnamefont {N.~B.}\ \bibnamefont {Perkins}},\
  and\ \bibinfo {author} {\bibfnamefont {G.~B.}\ \bibnamefont {Hal\'asz}},\
  }\bibfield  {title} {\bibinfo {title} {{Vacancy Spectroscopy of Non-Abelian
  Kitaev Spin Liquids}},\ }\href
  {https://doi.org/10.1103/PhysRevLett.132.136503} {\bibfield  {journal}
  {\bibinfo  {journal} {Physical Review Letters}\ }\textbf {\bibinfo {volume}
  {132}},\ \bibinfo {pages} {136503} (\bibinfo {year}
  {2024}{\natexlab{a}})}\BibitemShut {NoStop}%
\bibitem [{\citenamefont {Kao}\ \emph {et~al.}(2024{\natexlab{b}})\citenamefont
  {Kao}, \citenamefont {Hal\'asz},\ and\ \citenamefont {Perkins}}]{Kao2024b}%
  \BibitemOpen
  \bibfield  {author} {\bibinfo {author} {\bibfnamefont {W.-H.}\ \bibnamefont
  {Kao}}, \bibinfo {author} {\bibfnamefont {G.~B.}\ \bibnamefont {Hal\'asz}},\
  and\ \bibinfo {author} {\bibfnamefont {N.~B.}\ \bibnamefont {Perkins}},\
  }\bibfield  {title} {\bibinfo {title} {{Dynamics of Vacancy-Induced Modes in
  the Non-Abelian Kitaev Spin Liquid}},\ }\href
  {https://doi.org/10.1103/PhysRevB.109.125150} {\bibfield  {journal} {\bibinfo
   {journal} {Physical Review B}\ }\textbf {\bibinfo {volume} {109}},\ \bibinfo
  {pages} {125150} (\bibinfo {year} {2024}{\natexlab{b}})}\BibitemShut
  {NoStop}%
\bibitem [{\citenamefont {Takahashi}\ \emph {et~al.}(2023)\citenamefont
  {Takahashi}, \citenamefont {Yamada}, \citenamefont {Udagawa}, \citenamefont
  {Mizushima},\ and\ \citenamefont {Fujimoto}}]{Takahashi2023}%
  \BibitemOpen
  \bibfield  {author} {\bibinfo {author} {\bibfnamefont {M.~O.}\ \bibnamefont
  {Takahashi}}, \bibinfo {author} {\bibfnamefont {M.~G.}\ \bibnamefont
  {Yamada}}, \bibinfo {author} {\bibfnamefont {M.}~\bibnamefont {Udagawa}},
  \bibinfo {author} {\bibfnamefont {T.}~\bibnamefont {Mizushima}},\ and\
  \bibinfo {author} {\bibfnamefont {S.}~\bibnamefont {Fujimoto}},\ }\bibfield
  {title} {\bibinfo {title} {{Nonlocal Spin Correlation as a Signature of Ising
  Anyons Trapped in Vacancies of the Kitaev Spin Liquid}},\ }\href
  {https://doi.org/10.1103/PhysRevLett.131.236701} {\bibfield  {journal}
  {\bibinfo  {journal} {Phys. Rev. Lett.}\ }\textbf {\bibinfo {volume} {131}},\
  \bibinfo {pages} {236701} (\bibinfo {year} {2023})}\BibitemShut {NoStop}%
\bibitem [{\citenamefont {Imamura}\ \emph {et~al.}(2024)\citenamefont
  {Imamura}, \citenamefont {Mizukami}, \citenamefont {Tanaka}, \citenamefont
  {Grasset}, \citenamefont {Konczykowski}, \citenamefont {Kurita},
  \citenamefont {Tanaka}, \citenamefont {Matsuda}, \citenamefont {Yamada},
  \citenamefont {Hashimoto},\ and\ \citenamefont {Shibauchi}}]{Imamura2024}%
  \BibitemOpen
  \bibfield  {author} {\bibinfo {author} {\bibfnamefont {K.}~\bibnamefont
  {Imamura}}, \bibinfo {author} {\bibfnamefont {Y.}~\bibnamefont {Mizukami}},
  \bibinfo {author} {\bibfnamefont {O.}~\bibnamefont {Tanaka}}, \bibinfo
  {author} {\bibfnamefont {R.}~\bibnamefont {Grasset}}, \bibinfo {author}
  {\bibfnamefont {M.}~\bibnamefont {Konczykowski}}, \bibinfo {author}
  {\bibfnamefont {N.}~\bibnamefont {Kurita}}, \bibinfo {author} {\bibfnamefont
  {H.}~\bibnamefont {Tanaka}}, \bibinfo {author} {\bibfnamefont
  {Y.}~\bibnamefont {Matsuda}}, \bibinfo {author} {\bibfnamefont {M.~G.}\
  \bibnamefont {Yamada}}, \bibinfo {author} {\bibfnamefont {K.}~\bibnamefont
  {Hashimoto}},\ and\ \bibinfo {author} {\bibfnamefont {T.}~\bibnamefont
  {Shibauchi}},\ }\bibfield  {title} {\bibinfo {title} {{Defect-Induced
  Low-Energy Majorana Excitations in the Kitaev Magnet
  $\ensuremath{\alpha}\text{\ensuremath{-}}{\mathrm{RuCl}}_{3}$}},\ }\href
  {https://doi.org/10.1103/PhysRevX.14.011045} {\bibfield  {journal} {\bibinfo
  {journal} {Phys. Rev. X}\ }\textbf {\bibinfo {volume} {14}},\ \bibinfo
  {pages} {011045} (\bibinfo {year} {2024})}\BibitemShut {NoStop}%
\bibitem [{\citenamefont {Yatsuta}\ and\ \citenamefont
  {Mross}(2024)}]{Yatsuta2024}%
  \BibitemOpen
  \bibfield  {author} {\bibinfo {author} {\bibfnamefont {I.}~\bibnamefont
  {Yatsuta}}\ and\ \bibinfo {author} {\bibfnamefont {D.~F.}\ \bibnamefont
  {Mross}},\ }\bibfield  {title} {\bibinfo {title} {{Vacancies in Generic
  Kitaev Spin Liquids}},\ }\href
  {https://doi.org/10.1103/PhysRevLett.133.226501} {\bibfield  {journal}
  {\bibinfo  {journal} {Phys. Rev. Lett.}\ }\textbf {\bibinfo {volume} {133}},\
  \bibinfo {pages} {226501} (\bibinfo {year} {2024})}\BibitemShut {NoStop}%
\bibitem [{\citenamefont {Dantas}\ \emph {et~al.}(2024)\citenamefont {Dantas},
  \citenamefont {Kao},\ and\ \citenamefont {Perkins}}]{Dantas2024}%
  \BibitemOpen
  \bibfield  {author} {\bibinfo {author} {\bibfnamefont {V.}~\bibnamefont
  {Dantas}}, \bibinfo {author} {\bibfnamefont {W.-H.}\ \bibnamefont {Kao}},\
  and\ \bibinfo {author} {\bibfnamefont {N.~B.}\ \bibnamefont {Perkins}},\
  }\bibfield  {title} {\bibinfo {title} {{Phonon dynamics in the
  site-disordered Kitaev spin liquid}},\ }\href
  {https://doi.org/10.1103/PhysRevB.110.104425} {\bibfield  {journal} {\bibinfo
   {journal} {Phys. Rev. B}\ }\textbf {\bibinfo {volume} {110}},\ \bibinfo
  {pages} {104425} (\bibinfo {year} {2024})}\BibitemShut {NoStop}%
\bibitem [{\citenamefont {K{\"o}nig}\ \emph {et~al.}(2020)\citenamefont
  {K{\"o}nig}, \citenamefont {Randeria},\ and\ \citenamefont
  {J{\"a}ck}}]{konig2020tunneling}%
  \BibitemOpen
  \bibfield  {author} {\bibinfo {author} {\bibfnamefont {E.~J.}\ \bibnamefont
  {K{\"o}nig}}, \bibinfo {author} {\bibfnamefont {M.~T.}\ \bibnamefont
  {Randeria}},\ and\ \bibinfo {author} {\bibfnamefont {B.}~\bibnamefont
  {J{\"a}ck}},\ }\bibfield  {title} {\bibinfo {title} {{Tunneling Spectroscopy
  of Quantum Spin Liquids}},\ }\href
  {https://doi.org/10.1103/PhysRevLett.125.267206} {\bibfield  {journal}
  {\bibinfo  {journal} {Phys. Rev. Lett.}\ }\textbf {\bibinfo {volume} {125}},\
  \bibinfo {pages} {267206} (\bibinfo {year} {2020})}\BibitemShut {NoStop}%
\bibitem [{\citenamefont {Feldmeier}\ \emph {et~al.}(2020)\citenamefont
  {Feldmeier}, \citenamefont {Natori}, \citenamefont {Knap},\ and\
  \citenamefont {Knolle}}]{Feldmeier2020}%
  \BibitemOpen
  \bibfield  {author} {\bibinfo {author} {\bibfnamefont {J.}~\bibnamefont
  {Feldmeier}}, \bibinfo {author} {\bibfnamefont {W.}~\bibnamefont {Natori}},
  \bibinfo {author} {\bibfnamefont {M.}~\bibnamefont {Knap}},\ and\ \bibinfo
  {author} {\bibfnamefont {J.}~\bibnamefont {Knolle}},\ }\bibfield  {title}
  {\bibinfo {title} {{Local probes for charge-neutral edge states in
  two-dimensional quantum magnets}},\ }\href
  {https://doi.org/10.1103/PhysRevB.102.134423} {\bibfield  {journal} {\bibinfo
   {journal} {Phys. Rev. B}\ }\textbf {\bibinfo {volume} {102}},\ \bibinfo
  {pages} {134423} (\bibinfo {year} {2020})}\BibitemShut {NoStop}%
\bibitem [{\citenamefont {Udagawa}\ \emph {et~al.}(2021)\citenamefont
  {Udagawa}, \citenamefont {Takayoshi},\ and\ \citenamefont
  {Oka}}]{Udagawa2021}%
  \BibitemOpen
  \bibfield  {author} {\bibinfo {author} {\bibfnamefont {M.}~\bibnamefont
  {Udagawa}}, \bibinfo {author} {\bibfnamefont {S.}~\bibnamefont {Takayoshi}},\
  and\ \bibinfo {author} {\bibfnamefont {T.}~\bibnamefont {Oka}},\ }\bibfield
  {title} {\bibinfo {title} {{Scanning Tunneling Microscopy as a Single
  Majorana Detector of Kitaev's Chiral Spin Liquid}},\ }\href
  {https://doi.org/10.1103/PhysRevLett.126.127201} {\bibfield  {journal}
  {\bibinfo  {journal} {Phys. Rev. Lett.}\ }\textbf {\bibinfo {volume} {126}},\
  \bibinfo {pages} {127201} (\bibinfo {year} {2021})}\BibitemShut {NoStop}%
\bibitem [{\citenamefont {Bauer}\ \emph {et~al.}(2023)\citenamefont {Bauer},
  \citenamefont {Freitas}, \citenamefont {Pereira},\ and\ \citenamefont
  {Egger}}]{Bauer2023}%
  \BibitemOpen
  \bibfield  {author} {\bibinfo {author} {\bibfnamefont {T.}~\bibnamefont
  {Bauer}}, \bibinfo {author} {\bibfnamefont {L.~R.~D.}\ \bibnamefont
  {Freitas}}, \bibinfo {author} {\bibfnamefont {R.~G.}\ \bibnamefont
  {Pereira}},\ and\ \bibinfo {author} {\bibfnamefont {R.}~\bibnamefont
  {Egger}},\ }\bibfield  {title} {\bibinfo {title} {{Scanning Tunneling
  Spectroscopy of Majorana Zero Modes in a Kitaev Spin Liquid}},\ }\href
  {https://doi.org/10.1103/PhysRevB.107.054432} {\bibfield  {journal} {\bibinfo
   {journal} {Physical Review B}\ }\textbf {\bibinfo {volume} {107}},\ \bibinfo
  {pages} {054432} (\bibinfo {year} {2023})}\BibitemShut {NoStop}%
\bibitem [{\citenamefont {Peri}\ \emph {et~al.}(2024)\citenamefont {Peri},
  \citenamefont {Ilani}, \citenamefont {Lee},\ and\ \citenamefont
  {Refael}}]{Peri2024}%
  \BibitemOpen
  \bibfield  {author} {\bibinfo {author} {\bibfnamefont {V.}~\bibnamefont
  {Peri}}, \bibinfo {author} {\bibfnamefont {S.}~\bibnamefont {Ilani}},
  \bibinfo {author} {\bibfnamefont {P.~A.}\ \bibnamefont {Lee}},\ and\ \bibinfo
  {author} {\bibfnamefont {G.}~\bibnamefont {Refael}},\ }\bibfield  {title}
  {\bibinfo {title} {{Probing Quantum Spin Liquids with a Quantum Twisting
  Microscope}},\ }\href {https://doi.org/10.1103/PhysRevB.109.035127}
  {\bibfield  {journal} {\bibinfo  {journal} {Physical Review B}\ }\textbf
  {\bibinfo {volume} {109}},\ \bibinfo {pages} {035127} (\bibinfo {year}
  {2024})}\BibitemShut {NoStop}%
\bibitem [{\citenamefont {Kohsaka}\ \emph {et~al.}(2024)\citenamefont
  {Kohsaka}, \citenamefont {Akutagawa}, \citenamefont {Omachi}, \citenamefont
  {Iwamichi}, \citenamefont {Ono}, \citenamefont {Tanaka}, \citenamefont
  {Tateishi}, \citenamefont {Murayama}, \citenamefont {Suetsugu}, \citenamefont
  {Hashimoto}, \citenamefont {Shibauchi}, \citenamefont {Takahashi},
  \citenamefont {Nikolaev}, \citenamefont {Mizushima}, \citenamefont
  {Fujimoto}, \citenamefont {Terashima}, \citenamefont {Asaba}, \citenamefont
  {Kasahara},\ and\ \citenamefont {Matsuda}}]{Kohsaka2024}%
  \BibitemOpen
  \bibfield  {author} {\bibinfo {author} {\bibfnamefont {Y.}~\bibnamefont
  {Kohsaka}}, \bibinfo {author} {\bibfnamefont {S.}~\bibnamefont {Akutagawa}},
  \bibinfo {author} {\bibfnamefont {S.}~\bibnamefont {Omachi}}, \bibinfo
  {author} {\bibfnamefont {Y.}~\bibnamefont {Iwamichi}}, \bibinfo {author}
  {\bibfnamefont {T.}~\bibnamefont {Ono}}, \bibinfo {author} {\bibfnamefont
  {I.}~\bibnamefont {Tanaka}}, \bibinfo {author} {\bibfnamefont
  {S.}~\bibnamefont {Tateishi}}, \bibinfo {author} {\bibfnamefont
  {H.}~\bibnamefont {Murayama}}, \bibinfo {author} {\bibfnamefont
  {S.}~\bibnamefont {Suetsugu}}, \bibinfo {author} {\bibfnamefont
  {K.}~\bibnamefont {Hashimoto}}, \bibinfo {author} {\bibfnamefont
  {T.}~\bibnamefont {Shibauchi}}, \bibinfo {author} {\bibfnamefont {M.~O.}\
  \bibnamefont {Takahashi}}, \bibinfo {author} {\bibfnamefont {S.}~\bibnamefont
  {Nikolaev}}, \bibinfo {author} {\bibfnamefont {T.}~\bibnamefont {Mizushima}},
  \bibinfo {author} {\bibfnamefont {S.}~\bibnamefont {Fujimoto}}, \bibinfo
  {author} {\bibfnamefont {T.}~\bibnamefont {Terashima}}, \bibinfo {author}
  {\bibfnamefont {T.}~\bibnamefont {Asaba}}, \bibinfo {author} {\bibfnamefont
  {Y.}~\bibnamefont {Kasahara}},\ and\ \bibinfo {author} {\bibfnamefont
  {Y.}~\bibnamefont {Matsuda}},\ }\bibfield  {title} {\bibinfo {title}
  {{Imaging Quantum Interference in a Monolayer Kitaev Quantum Spin Liquid
  Candidate}},\ }\href {https://doi.org/10.1103/PhysRevX.14.041026} {\bibfield
  {journal} {\bibinfo  {journal} {Phys. Rev. X}\ }\textbf {\bibinfo {volume}
  {14}},\ \bibinfo {pages} {041026} (\bibinfo {year} {2024})}\BibitemShut
  {NoStop}%
\bibitem [{\citenamefont {Bauer}\ \emph {et~al.}(2024)\citenamefont {Bauer},
  \citenamefont {Freitas}, \citenamefont {Andrade}, \citenamefont {Egger},\
  and\ \citenamefont {Pereira}}]{Bauer2024}%
  \BibitemOpen
  \bibfield  {author} {\bibinfo {author} {\bibfnamefont {T.}~\bibnamefont
  {Bauer}}, \bibinfo {author} {\bibfnamefont {L.~R.~D.}\ \bibnamefont
  {Freitas}}, \bibinfo {author} {\bibfnamefont {E.~C.}\ \bibnamefont
  {Andrade}}, \bibinfo {author} {\bibfnamefont {R.}~\bibnamefont {Egger}},\
  and\ \bibinfo {author} {\bibfnamefont {R.~G.}\ \bibnamefont {Pereira}},\
  }\bibfield  {title} {\bibinfo {title} {{Local spin-flip transitions induced
  by magnetic quantum impurities in two-dimensional magnets}},\ }\href
  {https://doi.org/10.1103/PhysRevB.110.L220403} {\bibfield  {journal}
  {\bibinfo  {journal} {Phys. Rev. B}\ }\textbf {\bibinfo {volume} {110}},\
  \bibinfo {pages} {L220403} (\bibinfo {year} {2024})}\BibitemShut {NoStop}%
\bibitem [{\citenamefont {Jahin}\ \emph {et~al.}(2025)\citenamefont {Jahin},
  \citenamefont {Zhang}, \citenamefont {Hal\'asz},\ and\ \citenamefont
  {Lin}}]{Jahin2025}%
  \BibitemOpen
  \bibfield  {author} {\bibinfo {author} {\bibfnamefont {A.}~\bibnamefont
  {Jahin}}, \bibinfo {author} {\bibfnamefont {H.}~\bibnamefont {Zhang}},
  \bibinfo {author} {\bibfnamefont {G.~B.}\ \bibnamefont {Hal\'asz}},\ and\
  \bibinfo {author} {\bibfnamefont {S.-Z.}\ \bibnamefont {Lin}},\ }\bibfield
  {title} {\bibinfo {title} {{Quasiparticle Interference in Kitaev Quantum Spin
  Liquids}},\ }\href {https://doi.org/10.1103/PhysRevLett.134.126501}
  {\bibfield  {journal} {\bibinfo  {journal} {Phys. Rev. Lett.}\ }\textbf
  {\bibinfo {volume} {134}},\ \bibinfo {pages} {126501} (\bibinfo {year}
  {2025})}\BibitemShut {NoStop}%
\bibitem [{\citenamefont {Zhang}\ \emph
  {et~al.}(2025{\natexlab{a}})\citenamefont {Zhang}, \citenamefont {Batista},\
  and\ \citenamefont {Hal\'asz}}]{Zhang2025PRB}%
  \BibitemOpen
  \bibfield  {author} {\bibinfo {author} {\bibfnamefont {S.-S.}\ \bibnamefont
  {Zhang}}, \bibinfo {author} {\bibfnamefont {C.~D.}\ \bibnamefont {Batista}},\
  and\ \bibinfo {author} {\bibfnamefont {G.~B.}\ \bibnamefont {Hal\'asz}},\
  }\bibfield  {title} {\bibinfo {title} {{Low-energy edge signatures of the
  Kitaev spin liquid}},\ }\href {https://doi.org/10.1103/PhysRevB.111.L161104}
  {\bibfield  {journal} {\bibinfo  {journal} {Phys. Rev. B}\ }\textbf {\bibinfo
  {volume} {111}},\ \bibinfo {pages} {L161104} (\bibinfo {year}
  {2025}{\natexlab{a}})}\BibitemShut {NoStop}%
\bibitem [{\citenamefont {Zhang}\ \emph
  {et~al.}(2025{\natexlab{b}})\citenamefont {Zhang}, \citenamefont
  {Hal{\'a}sz},\ and\ \citenamefont {Batista}}]{Zhang2025npj}%
  \BibitemOpen
  \bibfield  {author} {\bibinfo {author} {\bibfnamefont {S.-S.}\ \bibnamefont
  {Zhang}}, \bibinfo {author} {\bibfnamefont {G.~B.}\ \bibnamefont
  {Hal{\'a}sz}},\ and\ \bibinfo {author} {\bibfnamefont {C.~D.}\ \bibnamefont
  {Batista}},\ }\bibfield  {title} {\bibinfo {title} {{Probing chiral Kitaev
  spin liquids via dangling boundary fermions}},\ }\href
  {https://doi.org/10.1038/s41535-025-00770-7} {\bibfield  {journal} {\bibinfo
  {journal} {npj Quantum Materials}\ }\textbf {\bibinfo {volume} {10}},\
  \bibinfo {pages} {59} (\bibinfo {year} {2025}{\natexlab{b}})}\BibitemShut
  {NoStop}%
\bibitem [{\citenamefont {Klein}\ \emph {et~al.}(2018)\citenamefont {Klein},
  \citenamefont {MacNeill}, \citenamefont {Song}, \citenamefont {Larson},
  \citenamefont {Fang}, \citenamefont {Xu}, \citenamefont {Ribeiro},
  \citenamefont {Canfield}, \citenamefont {Kaxiras}, \citenamefont {Comin},\
  and\ \citenamefont {Jarillo-Herrero}}]{Klein2018}%
  \BibitemOpen
  \bibfield  {author} {\bibinfo {author} {\bibfnamefont {D.~R.}\ \bibnamefont
  {Klein}}, \bibinfo {author} {\bibfnamefont {D.}~\bibnamefont {MacNeill}},
  \bibinfo {author} {\bibfnamefont {Q.}~\bibnamefont {Song}}, \bibinfo {author}
  {\bibfnamefont {D.~T.}\ \bibnamefont {Larson}}, \bibinfo {author}
  {\bibfnamefont {S.}~\bibnamefont {Fang}}, \bibinfo {author} {\bibfnamefont
  {M.}~\bibnamefont {Xu}}, \bibinfo {author} {\bibfnamefont {R.~A.}\
  \bibnamefont {Ribeiro}}, \bibinfo {author} {\bibfnamefont {P.~C.}\
  \bibnamefont {Canfield}}, \bibinfo {author} {\bibfnamefont {E.}~\bibnamefont
  {Kaxiras}}, \bibinfo {author} {\bibfnamefont {R.}~\bibnamefont {Comin}},\
  and\ \bibinfo {author} {\bibfnamefont {P.}~\bibnamefont {Jarillo-Herrero}},\
  }\bibfield  {title} {\bibinfo {title} {{Probing magnetism in
  two‐dimensional van der Waals crystalline insulators via electron
  tunneling}},\ }\href@noop {} {\bibfield  {journal} {\bibinfo  {journal}
  {Science}\ }\textbf {\bibinfo {volume} {360}},\ \bibinfo {pages} {1218}
  (\bibinfo {year} {2018})}\BibitemShut {NoStop}%
\bibitem [{\citenamefont {Lieb}(1994)}]{Lieb1994}%
  \BibitemOpen
  \bibfield  {author} {\bibinfo {author} {\bibfnamefont {E.~H.}\ \bibnamefont
  {Lieb}},\ }\bibfield  {title} {\bibinfo {title} {{Flux Phase of the
  Half-Filled Band}},\ }\href {https://doi.org/10.1103/PhysRevLett.73.2158}
  {\bibfield  {journal} {\bibinfo  {journal} {Phys. Rev. Lett.}\ }\textbf
  {\bibinfo {volume} {73}},\ \bibinfo {pages} {2158} (\bibinfo {year}
  {1994})}\BibitemShut {NoStop}%
\bibitem [{\citenamefont {Zschocke}\ and\ \citenamefont
  {Vojta}(2015)}]{ZschockeVojta2015}%
  \BibitemOpen
  \bibfield  {author} {\bibinfo {author} {\bibfnamefont {F.}~\bibnamefont
  {Zschocke}}\ and\ \bibinfo {author} {\bibfnamefont {M.}~\bibnamefont
  {Vojta}},\ }\bibfield  {title} {\bibinfo {title} {{Physical states and
  finite-size effects in Kitaev's honeycomb model: Bond disorder, spin
  excitations, and NMR line shape}},\ }\href
  {https://doi.org/10.1103/PhysRevB.92.014403} {\bibfield  {journal} {\bibinfo
  {journal} {Phys. Rev. B}\ }\textbf {\bibinfo {volume} {92}},\ \bibinfo
  {pages} {014403} (\bibinfo {year} {2015})}\BibitemShut {NoStop}%
\bibitem [{\citenamefont {Vojta}\ \emph {et~al.}(2016)\citenamefont {Vojta},
  \citenamefont {Mitchell},\ and\ \citenamefont
  {Zschocke}}]{Vojta2016impurity}%
  \BibitemOpen
  \bibfield  {author} {\bibinfo {author} {\bibfnamefont {M.}~\bibnamefont
  {Vojta}}, \bibinfo {author} {\bibfnamefont {A.~K.}\ \bibnamefont
  {Mitchell}},\ and\ \bibinfo {author} {\bibfnamefont {F.}~\bibnamefont
  {Zschocke}},\ }\bibfield  {title} {\bibinfo {title} {Kondo impurities in the
  kitaev spin liquid: Numerical renormalization group solution and
  gauge-flux-driven screening},\ }\href
  {https://doi.org/10.1103/PhysRevLett.117.037202} {\bibfield  {journal}
  {\bibinfo  {journal} {Phys. Rev. Lett.}\ }\textbf {\bibinfo {volume} {117}},\
  \bibinfo {pages} {037202} (\bibinfo {year} {2016})}\BibitemShut {NoStop}%
\bibitem [{\citenamefont {Takahashi}\ \emph {et~al.}(2025)\citenamefont
  {Takahashi}, \citenamefont {Kao}, \citenamefont {Fujimoto},\ and\
  \citenamefont {Perkins}}]{Takahashi2025higherS}%
  \BibitemOpen
  \bibfield  {author} {\bibinfo {author} {\bibfnamefont {M.~O.}\ \bibnamefont
  {Takahashi}}, \bibinfo {author} {\bibfnamefont {W.-H.}\ \bibnamefont {Kao}},
  \bibinfo {author} {\bibfnamefont {S.}~\bibnamefont {Fujimoto}},\ and\
  \bibinfo {author} {\bibfnamefont {N.~B.}\ \bibnamefont {Perkins}},\
  }\bibfield  {title} {\bibinfo {title} {Z2 flux binding to higher-spin
  impurities in the kitaev spin liquid},\ }\href
  {https://doi.org/10.1038/s41535-025-00729-8} {\bibfield  {journal} {\bibinfo
  {journal} {npj Quantum Materials}\ }\textbf {\bibinfo {volume} {10}},\
  \bibinfo {pages} {14} (\bibinfo {year} {2025})}\BibitemShut {NoStop}%
\bibitem [{\citenamefont {Knolle}\ \emph {et~al.}(2014)\citenamefont {Knolle},
  \citenamefont {Kovrizhin}, \citenamefont {Chalker},\ and\ \citenamefont
  {Moessner}}]{Knolle2014}%
  \BibitemOpen
  \bibfield  {author} {\bibinfo {author} {\bibfnamefont {J.}~\bibnamefont
  {Knolle}}, \bibinfo {author} {\bibfnamefont {D.~L.}\ \bibnamefont
  {Kovrizhin}}, \bibinfo {author} {\bibfnamefont {J.~T.}\ \bibnamefont
  {Chalker}},\ and\ \bibinfo {author} {\bibfnamefont {R.}~\bibnamefont
  {Moessner}},\ }\bibfield  {title} {\bibinfo {title} {Dynamics of a
  two-dimensional quantum spin liquid: Signatures of emergent majorana fermions
  and fluxes},\ }\href {https://doi.org/10.1103/PhysRevLett.112.207203}
  {\bibfield  {journal} {\bibinfo  {journal} {Phys. Rev. Lett.}\ }\textbf
  {\bibinfo {volume} {112}},\ \bibinfo {pages} {207203} (\bibinfo {year}
  {2014})}\BibitemShut {NoStop}%
\bibitem [{\citenamefont {Knolle}\ \emph {et~al.}(2015)\citenamefont {Knolle},
  \citenamefont {Kovrizhin}, \citenamefont {Chalker},\ and\ \citenamefont
  {Moessner}}]{Knolle2015Fractionalization}%
  \BibitemOpen
  \bibfield  {author} {\bibinfo {author} {\bibfnamefont {J.}~\bibnamefont
  {Knolle}}, \bibinfo {author} {\bibfnamefont {D.~L.}\ \bibnamefont
  {Kovrizhin}}, \bibinfo {author} {\bibfnamefont {J.~T.}\ \bibnamefont
  {Chalker}},\ and\ \bibinfo {author} {\bibfnamefont {R.}~\bibnamefont
  {Moessner}},\ }\bibfield  {title} {\bibinfo {title} {{Dynamics of
  fractionalization in quantum spin liquids}},\ }\href
  {https://doi.org/10.1103/PhysRevB.92.115127} {\bibfield  {journal} {\bibinfo
  {journal} {Physical Review B}\ }\textbf {\bibinfo {volume} {92}},\ \bibinfo
  {pages} {115127} (\bibinfo {year} {2015})}\BibitemShut {NoStop}%
\bibitem [{\citenamefont {Panigrahi}\ \emph {et~al.}(2023)\citenamefont
  {Panigrahi}, \citenamefont {Coleman},\ and\ \citenamefont
  {Tsvelik}}]{Aaditya2023}%
  \BibitemOpen
  \bibfield  {author} {\bibinfo {author} {\bibfnamefont {A.}~\bibnamefont
  {Panigrahi}}, \bibinfo {author} {\bibfnamefont {P.}~\bibnamefont {Coleman}},\
  and\ \bibinfo {author} {\bibfnamefont {A.}~\bibnamefont {Tsvelik}},\
  }\bibfield  {title} {\bibinfo {title} {{Analytic calculation of the vison gap
  in the Kitaev spin liquid}},\ }\href
  {https://doi.org/10.1103/PhysRevB.108.045151} {\bibfield  {journal} {\bibinfo
   {journal} {Phys. Rev. B}\ }\textbf {\bibinfo {volume} {108}},\ \bibinfo
  {pages} {045151} (\bibinfo {year} {2023})}\BibitemShut {NoStop}%
\bibitem [{\citenamefont {Baskaran}\ \emph {et~al.}(2007)\citenamefont
  {Baskaran}, \citenamefont {Mandal},\ and\ \citenamefont
  {Shankar}}]{Baskaran2007}%
  \BibitemOpen
  \bibfield  {author} {\bibinfo {author} {\bibfnamefont {G.}~\bibnamefont
  {Baskaran}}, \bibinfo {author} {\bibfnamefont {S.}~\bibnamefont {Mandal}},\
  and\ \bibinfo {author} {\bibfnamefont {R.}~\bibnamefont {Shankar}},\
  }\bibfield  {title} {\bibinfo {title} {{Exact Results for Spin Dynamics and
  Fractionalization in the Kitaev Model}},\ }\href
  {https://doi.org/10.1103/PhysRevLett.98.247201} {\bibfield  {journal}
  {\bibinfo  {journal} {Phys. Rev. Lett.}\ }\textbf {\bibinfo {volume} {98}},\
  \bibinfo {pages} {247201} (\bibinfo {year} {2007})}\BibitemShut {NoStop}%
\bibitem [{not()}]{noteVacancies}%
  \BibitemOpen
  \href@noop {} {}\bibinfo {note} {A small amount of spectral weight in the
  shaded region can also originate from bulk contributions associated with
  configurations in which two vacancies are adjacent. Such configurations
  significantly reduce the local flux excitation energy and generate bulk
  spectral weight extending to very low energies. At fixed dilute vacancy
  density, the probability of these configurations decreases with increasing
  system size. Accordingly, they are excluded from the disorder realizations
  considered here, as they do not qualitatively affect the resulting spectra or
  the identification of the vacancy-induced low-energy features.}\BibitemShut
  {Stop}%
\bibitem [{\citenamefont {Pereira}\ \emph {et~al.}(2006)\citenamefont
  {Pereira}, \citenamefont {Guinea}, \citenamefont {Lopes~dos Santos},
  \citenamefont {Peres},\ and\ \citenamefont {Castro~Neto}}]{Pereira2006}%
  \BibitemOpen
  \bibfield  {author} {\bibinfo {author} {\bibfnamefont {V.~M.}\ \bibnamefont
  {Pereira}}, \bibinfo {author} {\bibfnamefont {F.}~\bibnamefont {Guinea}},
  \bibinfo {author} {\bibfnamefont {J.~M.~B.}\ \bibnamefont {Lopes~dos
  Santos}}, \bibinfo {author} {\bibfnamefont {N.~M.~R.}\ \bibnamefont
  {Peres}},\ and\ \bibinfo {author} {\bibfnamefont {A.~H.}\ \bibnamefont
  {Castro~Neto}},\ }\bibfield  {title} {\bibinfo {title} {Disorder induced
  localized states in graphene},\ }\href
  {https://doi.org/10.1103/PhysRevLett.96.036801} {\bibfield  {journal}
  {\bibinfo  {journal} {Phys. Rev. Lett.}\ }\textbf {\bibinfo {volume} {96}},\
  \bibinfo {pages} {036801} (\bibinfo {year} {2006})}\BibitemShut {NoStop}%
\bibitem [{\citenamefont {Pereira}\ \emph {et~al.}(2008)\citenamefont
  {Pereira}, \citenamefont {Lopes~dos Santos},\ and\ \citenamefont
  {Castro~Neto}}]{Pereira2008}%
  \BibitemOpen
  \bibfield  {author} {\bibinfo {author} {\bibfnamefont {V.~M.}\ \bibnamefont
  {Pereira}}, \bibinfo {author} {\bibfnamefont {J.~M.~B.}\ \bibnamefont
  {Lopes~dos Santos}},\ and\ \bibinfo {author} {\bibfnamefont {A.~H.}\
  \bibnamefont {Castro~Neto}},\ }\bibfield  {title} {\bibinfo {title} {Modeling
  disorder in graphene},\ }\href {https://doi.org/10.1103/PhysRevB.77.115109}
  {\bibfield  {journal} {\bibinfo  {journal} {Phys. Rev. B}\ }\textbf {\bibinfo
  {volume} {77}},\ \bibinfo {pages} {115109} (\bibinfo {year}
  {2008})}\BibitemShut {NoStop}%
\bibitem [{\citenamefont {Saremi}(2007)}]{Saremi2007_RKKY_Graphene}%
  \BibitemOpen
  \bibfield  {author} {\bibinfo {author} {\bibfnamefont {S.}~\bibnamefont
  {Saremi}},\ }\bibfield  {title} {\bibinfo {title} {Rkky in half-filled
  bipartite lattices: Graphene as an example},\ }\href
  {https://doi.org/10.1103/PhysRevB.76.184430} {\bibfield  {journal} {\bibinfo
  {journal} {Phys. Rev. B}\ }\textbf {\bibinfo {volume} {76}},\ \bibinfo
  {pages} {184430} (\bibinfo {year} {2007})}\BibitemShut {NoStop}%
\bibitem [{\citenamefont {Sherafati}\ and\ \citenamefont
  {Satpathy}(2011)}]{Sherafati2011_RKKY}%
  \BibitemOpen
  \bibfield  {author} {\bibinfo {author} {\bibfnamefont {M.}~\bibnamefont
  {Sherafati}}\ and\ \bibinfo {author} {\bibfnamefont {S.}~\bibnamefont
  {Satpathy}},\ }\bibfield  {title} {\bibinfo {title} {Rkky interaction in
  graphene from the lattice green's function},\ }\href
  {https://doi.org/10.1103/PhysRevB.83.165425} {\bibfield  {journal} {\bibinfo
  {journal} {Phys. Rev. B}\ }\textbf {\bibinfo {volume} {83}},\ \bibinfo
  {pages} {165425} (\bibinfo {year} {2011})}\BibitemShut {NoStop}%
\bibitem [{\citenamefont {Kogan}(2013)}]{Kogan2013_RKKY_Gapped_Doped_Graphene}%
  \BibitemOpen
  \bibfield  {author} {\bibinfo {author} {\bibfnamefont {E.}~\bibnamefont
  {Kogan}},\ }\bibfield  {title} {\bibinfo {title} {Rkky interaction in gapped
  or doped graphene},\ }\href {https://doi.org/10.4236/graphene.2013.21002}
  {\bibfield  {journal} {\bibinfo  {journal} {Graphene}\ }\textbf {\bibinfo
  {volume} {2}},\ \bibinfo {pages} {8} (\bibinfo {year} {2013})}\BibitemShut
  {NoStop}%
\end{thebibliography}%

\end{document}